\documentclass[12pt]{article}

\newcommand{\rrangle}{\rangle\!\rangle}
\usepackage{graphicx,color}
\usepackage{amsmath}

\numberwithin{equation}{section}
\allowdisplaybreaks

\begin{document}


\begin{titlepage}
\thispagestyle{empty}
\begin{flushleft}
\hfill hep-th/0409069 \\
UT-04-23\hfill September, 2004 \\
\end{flushleft}

\vskip 1.5 cm
\bigskip

\begin{center}
{\LARGE Cardy states, factorization and}\\
{\LARGE idempotency in closed string field theory}

\renewcommand{\thefootnote}{\fnsymbol{footnote}}

\vskip 2cm
{\large
Isao~Kishimoto\footnote{e-mail address:
 ikishimo@hep-th.phys.s.u-tokyo.ac.jp} and
Yutaka~Matsuo\footnote{e-mail address:
 matsuo@phys.s.u-tokyo.ac.jp}}
\\
{\it
\noindent{ \bigskip }\\
Department of Physics, Faculty of Science, University of Tokyo \\
Hongo 7-3-1, Bunkyo-ku, Tokyo 113-0033, Japan\\
\noindent{ \smallskip }\\
}
\vfill
\end{center}
\begin{abstract}
We show that boundary states in the generic 
on-shell background
satisfy a universal nonlinear equation of closed string
field theory. It generalizes our previous claim for
the flat background.  The origin of the equation is
factorization relation of boundary conformal field theory
which is always true as an axiom.
The equation necessarily incorporates the information of
open string sector through a regularization, which
implies the equivalence with Cardy condition.
We also give a more direct proof
by oscillator representations for some nontrivial
backgrounds (torus and orbifolds).
Finally we discuss some properties of the closed string star product
for non-vanishing $B$ field and find that a
commutative and non-associative product (Strachan product)
appears naturally in Seiberg-Witten limit.
\end{abstract}\vfill
\end{titlepage}\setcounter{footnote}{0}
\renewcommand{\thefootnote}{\arabic{footnote}}

\newpage


\section{Introduction}
Since its discovery, D-brane has been one of the central
objects of interest in string theory. It represents
fundamental nonperturbative features and is an analog of
the soliton excitation in string theory.

In conformal field theory, the D-brane is described by
boundary state.  It belongs to the {\em closed} string Hilbert space
and is implemented by the boundary conditions such as,
\begin{equation}
 \partial_\tau X^\mu|_{\tau=0} |B\rangle_{\mathrm{Neumann}} = 0\,,\quad
 \mbox{or }\quad \partial_\sigma X^\mu|_{\tau=0} 
|B\rangle_{\mathrm{Dirichlet}} = 0\,.
\end{equation}
These equations  determine the state $|B\rangle$
up to normalization constant.  
The information of the open strings which live
on D-brane can be extracted from $|B\rangle$ 
after modular transformation,
\begin{equation}
 \langle B|q^{{1\over 2}\left(L_0+\tilde L_0-\frac{c}{12}\right)}
|B\rangle = 
\mbox{Tr}_{\mathcal{H}_{\mathrm{open}}} \tilde q^{L_0-\frac{c}{24}}\,,\quad
\tilde q\equiv e^{4\pi^2/\log q}\,.
\end{equation}

For more generic (conformal invariant)
background where we can not use the
free field oscillators as above, the boundary condition can be implemented
only through generators of  Virasoro algebra,
\begin{equation}\label{e_boundary}
 (L_n-\tilde L_{-n})|B\rangle =0\,.
\end{equation}
This condition is {\em universal} in a sense that
it does not depend on a particular representation
of Virasoro algebra
which corresponds to the background.

This linear equation, however, is not enough to characterize the 
D-brane completely.  We need further constraints that
 the open string sectors
derived from them should be well-defined.  More explicitly,
take two states $|B_i\rangle$ ($i=1,2$) 
which satisfy eq.~(\ref{e_boundary}).  The open string
sector appears in the annulus amplitude 
after the modular transformation can be written as,
\begin{equation}\label{e_Cardy}
  \chi_{12}(q) = \langle B_1 | 
q^{{1\over 2}\left(L_0+\tilde L_0-\frac{c}{12}\right)} |B_2\rangle
   = \sum_i \mathcal{N}_{12}^i\,
\chi_i(\tilde q)\,,\quad
\end{equation}
where $\chi_i$ are characters of the irreducible representations
in the open string channel.  In order to have well-defined
open string sector, the coefficients $\mathcal{N}_{12}^i$ must be non-negative
integers.  This is called {\em Cardy condition} \cite{Cardy:ir}.  We note that
these are non-linear (quadratic) constraints 
in terms of the boundary states.

These conditions (\ref{e_boundary}, \ref{e_Cardy}) are
written in terms of the boundary conformal field theory.
In this sense, it is at the level of the first quantization.
In order to consider the off-shell process, such as
tachyon condensation, we need to use the second quantized description.
One of the strong candidates of the off-shell descriptions
of string theory is string field theory.  Therefore, it is
natural to ask whether one may derive conditions which
are equivalent to (\ref{e_boundary}, \ref{e_Cardy}) in that language.

There are a few species of string field theories.  
The best-established one is Witten's
open bosonic string field theory \cite{Witten:1985cc}.
In terms of this formulation,
the annihilation process of unstable D-branes was first
studied extensively and it established the idea of ``tachyon vacuum''
by computation of the D-brane tension numerically \cite{Taylor:2003gn}.

In this paper, however, we do not use this formulation since the dynamical
variables of open string field theory depend essentially on the D-brane
where the open string is attached.  Since our goal
is to find the characterization of generic consistent boundaries,
this variable is not particularly natural because of this 
particular reference to the specific D-brane.

Since the boundary state belongs to the closed string Hilbert
space, we will instead take the closed string field theory
as the basic language.
With closed string variable, the description of the linear constraint
(\ref{e_boundary}) is trivial: $(L_n-\tilde L_{-n})\Phi=0$.
On the other hand, the description of the Cardy condition
(\ref{e_Cardy}) is much more nontrivial since it is the requirement
to the {\em open} string channel which appears only
after the modular transformation.
Furthermore, it is a nonlinear relation.  If it is possible
to represent it in string field theory, one needs to use
the closed string star product to express nonlinear relations.

\begin{figure}[htbp]
	\begin{center}
	\scalebox{0.5}[0.5]{\includegraphics{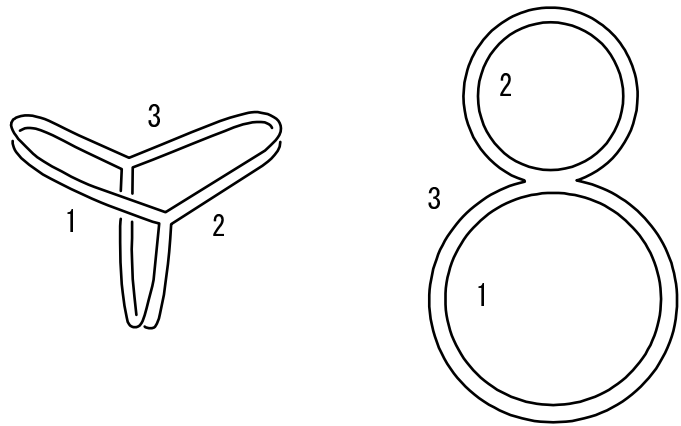}}
	\end{center}
	\caption{
Closed string vertices
}	\label{fig:ovelapping}
\end{figure}

There are two types of closed string star products which have been
studied in the literature.  The first one is Zwiebach's star product
which is a closed string version of Witten's open string star 
product \cite{Zwiebach, Saadi}
and the other one is a covariant version of the light-cone string
field theory (HIKKO's vertex) \cite{HIKKO2}. They are defined through
the overlap of three strings as depicted in Fig.~\ref{fig:ovelapping}.
These vertices are constructed to define the closed string field
theories proposed by these authors.  In this paper, however, our main
focus is the algebraic structure between boundary states.
Indeed, the nonlinear relation which we are going to study
holds for both of these two vertices.  In a sense, our relation
does not seem to be a 
consequence of their proposed action at least at this moment
but rather comes directly from the basic properties of the 
boundary conformal field theory.

The nonlinear equation has been proposed
and studied in our previous papers \cite{kmw1,kmw2,km1}.
It can be written as an idempotency relation\footnote{
This equation takes the form of the projector equation
of $C^*$ algebra. It reminds us of the fact that
the topological charge carried by the D-brane is given by the
K-theory \cite{r:K-theory}.  In the context of the noncommutative geometry,
an element which represents a K-theory class is given 
by the projection equation \cite{NCK},
\begin{equation}\label{e_projector}
 \phi \star \phi =\phi\,,
\end{equation}
where $\star$ is the product of the (noncommutative)
background geometry.  Solutions of this equation are called
noncommutative soliton \cite{Gopakumar:2000zd}.
Later, even in the bosonic
string which does not have RR charge, it was argued
\cite{Harvey:2000jt}
that the noncommutative solitons still represent
unstable D-branes.  Topological charges of D-branes
are represented in terms of the projector $\phi$.  For instance,
the D-brane number is related to the rank of $\phi$.
}
among boundary states,
\begin{equation}
\label{e_idempotency}
 \Phi \star \Phi = \mathcal{C}\,T_B^{-1}c_0^+ \Phi,\quad
 (\Phi\equiv c_0^-b_0^+|B\rangle\,,\,T_B\equiv \langle 0|
c_{-1}\tilde{c}_{-1}c_0^-|B\rangle)\,,
\end{equation}
where $T_B$ 
is the tension of the D-brane associated with
the boundary state $|B\rangle$.
In the first paper \cite{kmw1}, we proved it for the usual
D$p$-brane boundary states for $\Phi$ and HIKKO vertex for $\star$.
A surprise was that this equation is universal for
any boundary states which we considered including
the coefficient $\mathcal{C}$ \cite{kmw2}.
The proof is based on an explicit calculation with the
oscillator representation \cite{HIKKO2}.
In the second paper \cite{kmw2}, we gave an outline of the proof
of the same equation for Zwiebach's vertex.
The equation takes the same form for these two
vertices except for the overall constant $\mathcal{C}$.
{}From these observations, we conjectured that eq.~(\ref{e_idempotency})
is a background
independent  characterization of the boundary state.

Toward that direction, in \cite{kmw2}, we used the 
path integral definition of the string field theory 
in terms of the conformal mapping
\cite{LPP1} and tried to prove the relation in the general
background.  After some efforts, we have arrived at 
a weaker statement: suppose
$|B_i\rangle$ ($i=1,2$) satisfy the linear constraint
(\ref{e_boundary}), the state $|B_1\rangle \star |B_2\rangle $
also satisfies the same constraint.  It proves that product of
any boundary state in weak sense becomes again boundary
state in weak sense. This does not, of course, imply
the Cardy constraint (\ref{e_Cardy})
and, in particular, we could not understand the role of the
open string sector.

One of the purposes of this paper is to discuss the link
with open string which was missed in our previous studies
and establish more explicit relation between (\ref{e_idempotency}) and
the Cardy condition. A crucial hint to this problem 
is that the coefficient $\mathcal{C}$
in the relation (\ref{e_idempotency}) is actually divergent
and it is necessary to introduce some sort of regularization
in the computation.  In \cite{kmw1}, 
we cut-off the rank of Neumann coefficient by $K$. Then
the divergent coefficient behaves as $\mathcal{C}\sim K^3$.
In LPP approach, on the other hand, another
type of the regularization can be introduced
by slightly shifting the interaction point on the world sheet.
Such a shift gives a small strip which interpolates between
two holes associated with the boundary.
In the limit of turning-off the regulator, the moduli parameter
which describes the shape of the strip becomes zero.
This is the usual factorization process where the world sheet becomes
degenerate \cite{r:Factorization}.  
An essential point is that such a factorization process
occurs in the {\em open string channel} between the two holes.
In this way, the equation for the string field  (\ref{e_idempotency})
can be related with the consistency of the dual open string
channel.  The leading singularity comes from the propagation
of the open string tachyon which is universal for any boundary
states in arbitrary background and it explains our claim
that the divergent factor
$\mathcal{C}$ is also universal for any Cardy  states.

This outline will be explained in detail in section 2.
We will also repeat our previous discussion \cite{kmw2} that
only the boundary states can satisfy the equation (\ref{e_boundary}).
By combining these ideas, it will be obvious that the nonlinear
equation (\ref{e_boundary}) plays an essential r\^ole to understand
D-branes in the context of string field theory.

At this point, it may be worth while to mention that our claims
are remarkably similar to the scenario 
conjectured in vacuum string field theory \cite{r:VSFT}.
The form of the equation is exactly the same except that the dynamical
variables and the star product are totally different.  
There is no nontrivial solution in the vicinity of $\Phi=0$ and the
non-vanishing solutions correspond to D-branes.  
In a sense, our equation is an explicit realization of
VSFT scenario in the dual closed string channel.
While the dynamical variables is closed string field,
the physical excitations around the boundary state
are on-shell open string mode \cite{kmw1, kmw2}.

Our discussion in section 2 is based on the path integral
and the argument becomes necessarily formal to some extent.  
In this sense, it is desirable to check the consistency of the argument
in the oscillator representation for some non-trivial
backgrounds.  Fortunately, there are explicit forms of
the three string vertex for (1) toroidal $T^d$ and
(2) orbifold $T^d/Z_2$ compactification 
\cite{HIKKO_torus,Itoh_Kunitomo}. 

In both cases, the three string vertex has some modifications
compared to that on the flat background ${\bf R}^d$ \cite{HIKKO2}.
One needs to include cocycle factor due to the existence of
winding mode and take into account the twisted sector
in orbifold case.
The cocycle factor is needed to keep Jacobi identity
for the $\star$ product of the closed string field theory.
In section 3, we perform 
explicit computation of $\star$ product among
the boundary (Ishibashi) states.
For torus case, there appear extra cocycle factors
in the algebra of Ishibashi states and
some care is needed to construct Cardy states
as idempotents of the algebra.
For the orbifold case, mixing between untwisted sector
and twisted sector is needed to describe the 
fractional D-branes.  The ratio of the coefficients
of the two sectors is given as the ratio of the determinants
of the Neumann matrices for the (un)twisted sectors.
We use various regularization methods to calculate them
explicitly.
This result is consistent with our previous arguments \cite{km1}.

Finally, the third issue which will be discussed in section 4
is to incorpolate the noncommutativity on the D-brane.
We have already seen in \cite{kmw1} that non-commutativity
on the world volume of the D-brane forces us to use the open string
metric to write down the on-shell conditions.
This is possible since the explicit form of the
boundary state is known for such cases.
On the other hand, for the noncommutativity in the transverse
directions, it is difficult to express in the language
of the boundary conformal field theory.
This is, however, an important set-up for
matrix models or noncommutative Yang-Mills theory.

A motivation toward this direction
is our previous study \cite{kmw2} where we have seen that
an analog of the noncommutative 
soliton arises in the commutative limit. We note that
the idempotency relation for the D$p$-brane takes the following
form in the matter sector,
\begin{equation}
 |B,x^\perp\rangle \star |B, y^\perp\rangle
 =\mathcal{C}_d\,\delta^{d-p-1}(x^\perp-y^\perp)\,|B,y^\perp\rangle\,\,,
\end{equation}
where $x^\perp,\,y^\perp$ are the coordinates in the transverse
directions.  In order to recover the universal relation
(\ref{e_idempotency}), we need to take a linear combination,
\begin{equation}
 |B\rangle_f\equiv  \int d^{d-p-1} x^{\perp}\, f(x^\perp) |B,x^\perp\rangle
\,\,.
\end{equation}
Eq.~(\ref{e_idempotency}) then implies $f^2(x^\perp)=f(x^\perp)$
which is the same as (\ref{e_projector})
in the commutative limit.  It is, therefore, tempting
to study what kind of modification will be necessary
in the presence of $B$ field.

In order to study it, we consider a particular deformation of
Ishibashi states which seems to be relevant to 
describe the noncommutativity along the transverse directions.  
We take the star product between
them and take Seiberg-Witten limit. The constraint for $f$ is
deformed to the following form,
\begin{equation}
 (f\diamond f)(x^\perp) = f(x^\perp)\,\,,\qquad
 \diamond\equiv \frac{\sin(\Lambda)}{\Lambda}\,,\quad
 \Lambda=\frac{1}{2}\overleftarrow{\partial_i}\theta^{ij}
 \overrightarrow{\partial_j}\,.
\end{equation}
This $\diamond$ product is commutative but breaks associativity.
It appeared in mathematical literature \cite{Strachan}
and is related to the loop corrections in noncommutative
super Yang-Mills theory \cite{start}.  
The appearance of such deformation seems to be natural 
since the star product of two boundaries is topologically 
equivalent to one loop from the open string viewpoint.

\section{Idempotency relation in generic background}
In this section, we prove the relation (\ref{e_idempotency})
in the generic background by using a sequence of conformal
maps.  Our proof depends only on
a generic property of the boundary conformal field theory ---
factorization --- which should be satisfied axiomatically in any BCFT.
Our discussion also shows a clear link between the Cardy condition
and the idempotency relation.

\subsection{$\star$ product and factorization}
\begin{figure}[htbp]
	\begin{center}
	\scalebox{0.55}[0.55]{\includegraphics{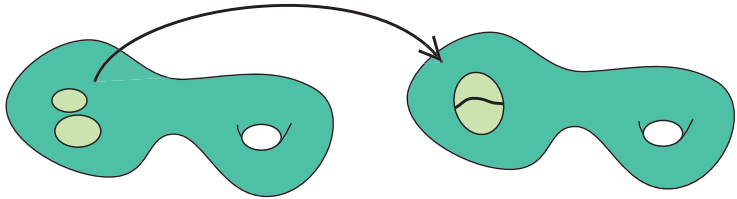}}
	\end{center}
	\caption{
Factorization associated with merging two holes.
}	\label{fig:degenerate}
\end{figure}

The factorization is a general behavior of the
correlation functions of the conformal
field theory defined on a pinched Riemann surface.
The relevant process for us is the degeneration of a strip
between two holes where
the correlation function behaves as
\begin{equation}
 \langle \mathcal{O}\cdots \rangle \rightarrow
  \sum_i \langle \mathcal{O}\cdots A_i(z_1) A_i(z_2)\rangle q^{\Delta_i}
\end{equation}
where $i$ is the label  of orthonormal basis $\left\{A_i(z)\right\}$
of the open string Hilbert space between two holes
and $\Delta_i$ is the conformal dimension of $A_i$.
The open string channel depends on the boundary conditions
at two holes.
$q$ is a real parameter which describes
the degeneration of the strip and $z_{1,2}$ are the  coordinates
of the two points along the boundary
where the two ends of the strip are attached (Fig.~\ref{fig:degenerate}).

\begin{figure}[htbp]
	\begin{center}
	\scalebox{0.5}[0.5]{\includegraphics{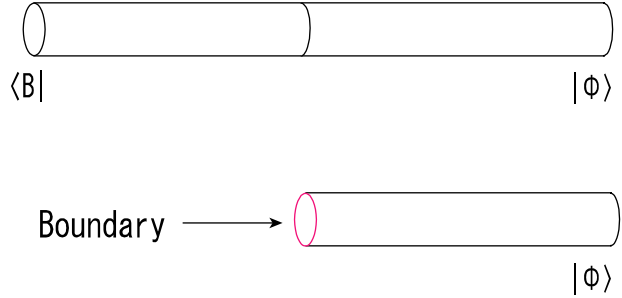}}
	\end{center}
	\caption{
Geometrical interpretation of boundary state as a surface state.
}	\label{fig:boundary}
\end{figure}

In the star product, such a  degeneration of a
strip appears
as we explained in our previous paper \cite{kmw2}.
It comes from combining the geometrical nature of
the boundary state as a surface state and three string vertex.
In order to explain the former, we consider an inner
product between a closed string state $|\chi\rangle$
and  a boundary state: $\langle B|\chi\rangle$.
On the world-sheet, it is equivalent to the one-point function
on a disk $\langle \chi(0) \rangle$ with the boundary condition
at $|z|=1$ specified by the boundary state.
If we map from the sphere to a cylinder by a conformal transformation,
 $w=\log z=\tau+i\sigma$, 
the geometrical role of the boundary state can be summarized as follows:
(Fig.~\ref{fig:boundary})
\begin{enumerate}
 \item cut the infinite cylinder at $\tau=0$ and strip off the region
       $\tau>0$,
 \item set the boundary condition specified by $|B\rangle$ at the boundary.
\end{enumerate}

\begin{figure}[htbp]
	\begin{center}
	\scalebox{0.9}[0.9]{\includegraphics{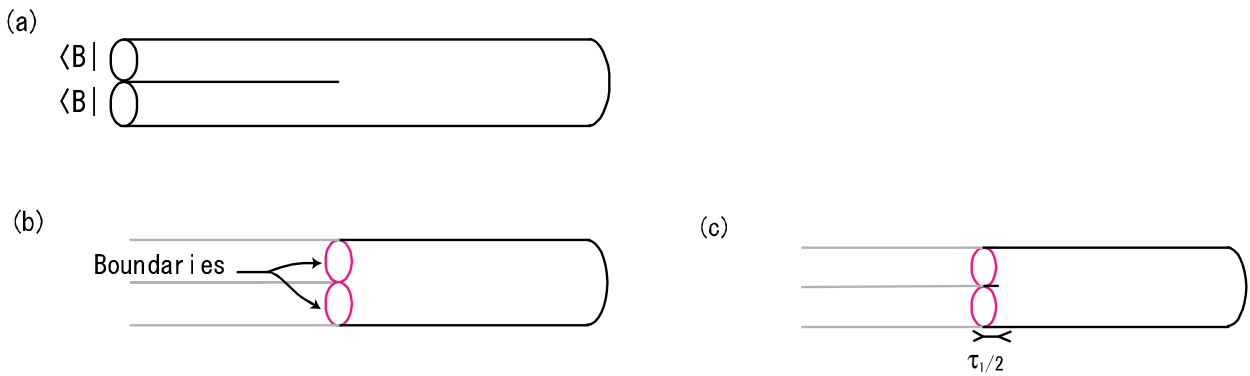}}
	\end{center}
	\caption{
(a) Putting boundary states at two legs of trousers that
is associated with three string vertex.
(b) Stripping two legs at the origin.
(c) Shifting the interaction point as a regularization.
}	\label{fig:vertex}
\end{figure}

We combine this property of the boundary state with
that of the three string vertex, which is
represented by Mandelstam diagram (Fig.~\ref{fig:vertex}-a)
for HIKKO type vertex.  
The matrix element $\Phi_1\cdot(\Phi_2\star\Phi_3)$ corresponds to
putting three local operators $\Phi_i$ ($i=1,2,3$) at the ends
of three half cylinders.

As we see previously, the boundary states are not 
described by local operators but should be interpreted as
the surface state. To take the product of
two boundary states $|B\rangle \star |B\rangle$ is then geometrically
represented as the Mandelstam diagram whose two legs are stripped
at the interaction time $\tau=0$ (see a Fig.~\ref{fig:vertex}-b).

This configuration is, however, singular since two boundaries are attached
at one point (interaction point) and we need a regularization 
to obtain a smooth surface. 
A natural regularization is to shift the location
of the boundary slightly, for example at $\tau=\tau_1/2>0$
(Fig.~\ref{fig:vertex}-c).
As we see later, this is equivalent to a cut-off of the
Neumann matrix with finite size $K$ 
which was used in our previous paper \cite{kmw1}. 
The correspondence of the regulator turns out to be
$K\sim \tau_1^{-1}$ .
With this regularization, the world-sheet becomes
a cylinder with one vertex operator insertion.
The limit $\tau_1\rightarrow 0$ is equivalent to
shrinking a strip of this diagram and reducing it to
a disk. We can use the discussion of factorization as,
\begin{equation}
 |B\rangle \star_{\tau_1} |B\rangle =
\sum_i q^{\Delta_i} A_i(\sigma_1) A_i(\sigma_2)|B\rangle\,,
\end{equation}
where again $A_i(\sigma_i)$ belongs to a set of orthonormal
operators in the {\em open} string Hilbert space
with both end attached to a brane specified by $|B\rangle$
and $\sigma_{1,2}$
are the coordinates along the boundary.

For the consistency of boundary states of the bosonic string,
the lowest dimensional operator in the Hilbert space
is always tachyon state which is written as,
$|0\rangle^{\mathrm{m}}\otimes c_1|0\rangle^{\mathrm{gh}}$
where $|0\rangle^{\mathrm{m,gh}}$ are $SL(2,R)$ invariant
vacuum for the matter and ghost.
The conformal dimension of this state is $-1$.
Other terms depend on the detail of the boundary state
but they always give less singular terms as $q\rightarrow 0$.
Similarly, if we $\star$-multiply two different boundary states
$|B_1\rangle \star |B_2\rangle$, the open string sector
is described by the Hilbert space of the {\em mixed} boundary condition
and the lowest dimensional operator always has a dimension $\Delta$
greater than $0$. This simple argument then implies,
\begin{equation}
 |B_\alpha\rangle \star |B_\beta\rangle \sim q^{-1} c(\sigma_1)
(c\partial c)(\sigma_2)   \delta_{\alpha\beta}|B_\beta\rangle + \ 
\mbox{less singular terms in $K$}\,.
\end{equation}
Although the less singular terms do depend on the background and
boundary state, the first term is universal.
As we see, the precise structure for ghost and singularity
is more involved due to the ghost insertions in the three string
vertex and the singular behaviour themselves should be modified
in order to obtain the precise agreement with the oscillator computation.

\subsection{Computation by conformal mappings
\label{sec:compC}
}

\begin{figure}[htbp]
	\begin{center}
\scalebox{0.8}[0.8]{\includegraphics{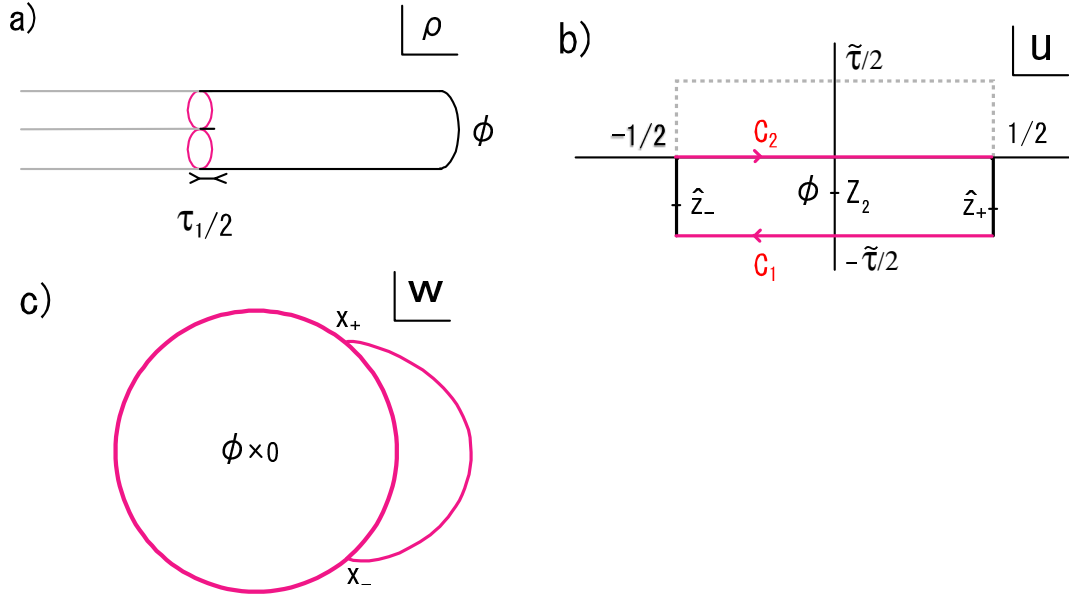}}
	\end{center}
	\caption{
(a) (reguralized) string vertex with two boundary states;
(b) a cylinder diagram;
(c) a disk diagram with operator insertions.
}	\label{fig:mapping}
\end{figure}

In order to see the degeneration in detail,
we consider three surfaces which
can be related with each other by conformal mappings
(Fig.~\ref{fig:mapping}-a,b,c).
The first one is the regularized version of
the Mandelstam diagram for
the star product of two boundary states (Fig.~\ref{fig:vertex}-a).
A natural coordinate for this diagram is  $\rho=\tau+i\sigma$
($\tau>0$,  $-\pi\alpha\leq \sigma\leq \pi\alpha$).
The interaction point is 
$\rho=\tau_1/2\pm i\pi\alpha_1$, 
($0\leq \alpha_1\leq \alpha$) and the parameter $\tau_1>0$ 
is introduced for the regularization.

This diagram has two holes together with one vertex operator insertion
at infinity.
Since it is topologically annulus, it can be
mapped to the standard annulus diagram (Fig.~\ref{fig:mapping}-b).
A complex parameter $u$ 
($|\mbox{Re}(u)|\leq 1/2$, $-t/2\leq\mbox{Im}(u)\leq 0$, $\tilde\tau:=it$) 
is a flat complex coordinate
and $\tilde\tau$ is the moduli parameter.
These two diagrams are related with each other
by a generalized Mandelstam mapping \cite{r:Mandel}\cite{r:AKT},
\begin{equation}
\label{eq:Mandel1loop}
 \rho(u) = \alpha \ln \frac{\vartheta_1(u-Z_1|\tilde\tau)}{\vartheta_1(u-Z_2|
\tilde\tau)}
- 2\pi i \alpha_1 u\,.
\end{equation}
We note that it can be extended as a mapping
between the doubles of above diagrams,
i.e., a torus $-\frac{1}{2}\leq \mbox{Re}(u)\leq \frac{1}{2}$,
$-t/2\leq \mbox{Im}(u)\leq t/2$ and corresponding Mandelstam
diagram. 
The parameters $Z_{1,2}:=\mp \beta\tilde\tau/2$ 
are mapped to the infinities $\mbox{Re}(\rho)=\mp \infty$. 
The interaction point 
$\tau_1/2 \mp i\pi\alpha_1$ 
is mapped to 
$\hat{z}_{\pm}=\pm \frac{1}{2}-\tilde{\tau}y$.
There is a set of relations among parameters \cite{r:AKT},
\begin{eqnarray}
 && \beta=-\frac{\alpha_1}{\alpha}\,,\\
 && \frac{\tau_1}{\alpha} =2 \ln 
\frac{\vartheta_2(\tilde\tau(-\beta/2+y)|\tilde\tau)}
{\vartheta_2(\tilde\tau(-\beta/2-y)|\tilde\tau)}
-4\pi i \beta\tilde\tau y\,,\\
&& g_2(\tilde\tau(-\beta/2+y)|\tilde\tau)+
g_2(\tilde\tau(-\beta/2-y)|\tilde\tau)=2\pi i \beta\,,
\end{eqnarray}
where $ g_2(\nu|\tilde\tau)=\partial_\nu \ln\vartheta_2(\nu|\tilde\tau)$.
In the degenerate limit $\tau_1\rightarrow 0$, these are
reduced to,
\begin{eqnarray}
 && y\sim\frac{1}{4}\,,\qquad
  e^{-\frac{i\pi}{\tilde\tau}}
\equiv q^{1/2}\sim\frac{\tau_1}{8\alpha\sin(-\pi\beta)}\,.
\label{eq:yq}
\end{eqnarray}

The third diagram (Fig.~\ref{fig:mapping}-c)
is disk-like diagram with two short slits.
It is parametrized by a complex coordinate $w$ with 
$|w|\leq 1$.  The relation with the Mandelstam diagram is 
very simple,
\begin{equation}
 w=f(\rho)\equiv \exp(-\rho/\alpha)\,.
\end{equation}
Two slits are located at $x_\pm=\exp(\pm \pi i\alpha_1/\alpha)
=\exp(\mp \pi i \beta)$.

Each diagram has its own role in the computation
of $|B\rangle \star |B\rangle$.  Firstly, 
the Mandelstam diagram gives the definition
of the star product. The expression,\footnote{
We use the notation $b_0^+=b_0+\tilde{b}_0,c_0^-=c_0-\tilde{c}_0$
and $b_0^-={1\over 2}(b_0-\tilde{b}_0),c_0^+={1\over
2}(c_0+\tilde{c}_0)$.
The extra ghost zero mode $c_0^-$ is needed for our convention of 
the HIKKO $\star$ product.
Here, we assign the string length parameter 
$\alpha_1,\,\alpha_2=\alpha-\alpha_1(>0)$ 
to each string $1,2$ in order to use the HIKKO 3-string vertex.
}
\begin{equation}
\mathcal{F}\equiv  \left(
\langle B_1|_{\tau_1\over 2\alpha_1}b^+_0c_0^-\star  
\langle B_2|_{\tau_1\over 2\alpha_2}b^+_0c_0^-  
\right)|\phi\rangle
 \quad (\langle B|_T\equiv 
\langle B|e^{-T(L_0+\bar L_0)}) \,,
\end{equation}
can be evaluated as the one point function of $\phi$ 
inserted at $\tau=\infty$ in $\rho$-plane with two boundaries defined by
$| B_{1,2}\rangle$.  
$b_0^+$ insertion is used to cancel
$c_0^+$ factor contained in the boundary state and to
set the ghost number to be two.  
The ket vector $|\phi\rangle =\phi(0)|0\rangle$ has ghost number
two as usual.
If we map it to the standard annulus diagram 
(Fig.~\ref{fig:mapping}-b),
it can be rewritten as,\footnote{
The prefactor $\alpha_1\alpha_2$ comes from the conformal factor 
$(dw_r/d\rho)^{-1}$ ($r=1,2$)
in gluing the local disks (in $w_r$-plane) 
which represents strings 1 and 2 to $\rho$-plane in
Fig.~\ref{fig:mapping}-a.
}
\begin{eqnarray}
\label{eq:calF}
\mathcal{F}&=&
\alpha_1\alpha_2
 \langle B_1 | \tilde{q}^{{1\over 2}(L_0+\tilde{L}_0)}\,b_\rho^+\, 
(\rho^{-1}\!\circ\phi)(Z_2)\, b_\rho^+ |B_2\rangle\,,\\
&&\tilde{q}:=e^{2\pi i\tilde{\tau}}\,,~~~~
b_\rho^+=\oint \frac{du}{2\pi i} b(u)
\left(\frac{d\rho}{du}\right)^{-1}+c.c.\,.
\end{eqnarray}

We need to evaluate this expression in the limit $\tilde\tau\rightarrow
0$ and take the conformal transformation to 
disk diagram
(Fig.~\ref{fig:mapping}-c). 

As we have argued, taking the limit 
corresponds to taking the lowest dimensional
operator in the {\em open} string channel.
Therefore, one can rewrite $\mathcal{F}$ in this limit as,
(after the conformal map to $w$ plane),
\begin{equation}
 \mathcal{F}\sim \delta_{12}\langle B|b_{w}^{1+}b_{w}^{2+}
 (g\circ c)(x_+) (g\circ c\partial c)(x_-)|\phi\rangle\,,
\end{equation}
where $x_{\pm}=e^{\mp i\pi\beta}$ are the locations of tachyon insertions
and $g=f\circ \rho$. $b^{1+}_w$ and $b^{2+}_w$
are the conformal transformations of 
$b_\rho^+$ along two boundaries.
The ghost insertions become very complicated but we have
already proved in the oscillator formulation \cite{kmw1}
that 
\begin{equation}
\label{eq:Bb1b2}
 \langle B|b_{w}^{1+}b_{w}^{2+}
 (g\circ c)(x_+) (g\circ c\partial c)(x_-)\sim\langle B|c_0^-\,.
\end{equation}
In order to fix the coefficient, we take the simplest example,
$|\phi\rangle=c_{1}\tilde{c}_{1}|0\rangle$
and calculate both sides of the equation.

It is convenient to divide the computation into matter and
ghost sectors.  Let us first consider the matter part.
The vertex operator for $|\phi\rangle$ is simply 1.
The inner product between two boundary states is
\begin{equation}
\langle B^{\mathrm{m}}_1|
{\tilde{q}}^{{1\over 2}\left(L_0+\tilde{L}_0-{c\over 12}\right)}
|B^\mathrm{m}_2\rangle =
q^{-{c\over 24}}\,\delta_{12}+\mbox{(higher order in $q$)}\,,
\label{eq:B1B2m}
\end{equation}
where we have supposed that these boundary states
$|B_1^{\rm m}\rangle,|B_2^{\rm m}\rangle $ satisfy Cardy condition
and used the leading behavior of the character of identity operator
 $\chi_1(q)\simeq q^{-{c\over 24}}$ ($q\rightarrow 0$).
On the other hand, the right hand side becomes,
\begin{equation}
 \langle B^\mathrm{m}_2|0\rangle =T_{B_2}\,.
\label{eq:TB2}
\end{equation}
Namely the tension for the D-brane \cite{Harvey:1999gq}.


For the ghost part,
we compute ${\cal F}$ (\ref{eq:calF}) with $\phi=c\tilde{c}$
and take the limit of ${\tilde{\tau}}\rightarrow +i0$ 
after modular transformation.
In order to compute explicitly, we map the $u$-plane to 
$\tilde{\rho}=2\pi i u$ 
such that $\tilde{\rho}$-plane becomes closed string strip 
with period $2\pi$ in the ${\rm Im}\tilde{\rho}$ direction and then
we expand the ghosts as
\begin{eqnarray}
&&b(\tilde{\rho})=\sum_{n=-\infty}^{\infty}b_ne^{-n\tilde{\rho}}\,,~~~
c(\tilde{\rho})=\sum_{n=-\infty}^{\infty}c_ne^{-n\tilde{\rho}}\,,~~~
\{b_n,c_m\}=\delta_{n+m,0}\,,~~~~~~
\end{eqnarray}
and similar ones for $\tilde{b}(\bar{\tilde{\rho}})$ and 
$\tilde{c}(\bar{\tilde{\rho}})$.
Using the property of the boundary states 
in the ghost sector: $(b_n-\tilde{b}_{-n})|B\rangle=0$, 
we calculate ghost contribution for ${\cal F}$ (\ref{eq:calF}) as
\begin{eqnarray}
{\cal F}_{c\tilde{c}}
&=&4\alpha_1\alpha_2(2\pi)^2
\int_{C_1} {du_1\over 2\pi i}
{du_1\over d\rho}
\int_{C_2} {du_2\over 2\pi i}{du_2\over d\rho}
\left[\left.{du\over dw_3}\right|_{w_3=0}
\left.{d\bar{u}\over d\bar{w}_3}\right|_{\bar{w}_3=0}
\right]^{-1}\nonumber\\
&&\times\,
\langle B|\tilde{q}^{{1\over 2}\left(L_0+\tilde{L}_0+{13\over 6}\right)}
b(2\pi i u_1)c(2\pi i Z_2)\tilde{c}(-2\pi i \bar{Z}_2)b(2\pi i u_2)
|B\rangle\,.~~~~~
\label{eq:Fcc}
\end{eqnarray}
($C_1$ and $C_2$ are given in Fig.~\ref{fig:mapping}-b.)
Here the conformal factor $C_{c\tilde{c}}$ for $c\tilde{c}$ (given by 
$[\cdots]^{-1}$ in the integrand) 
can be evaluated using the Mandelstam map (\ref{eq:Mandel1loop})
with $\rho(u)=\alpha_3 \log w_3$ ($\alpha_3=-\alpha_1-\alpha_2=-\alpha$)
for string $3$ region where $\phi$ is inserted:
\begin{eqnarray}
\label{eq:confC}
C_{c\tilde{c}}=
\left|{\vartheta_1(Z_2-Z_1|\tilde{\tau})\over 
\vartheta_1^{\prime}(0|\tilde{\tau})}e^{i\pi\tilde{\tau}\beta^2}
\right|^{-2}
\sim \pi^2|\tilde{\tau}|^{-2}(\sin\pi\beta)^{-2}\,,
~~~(\tilde{\tau}\rightarrow +i0)\,.
\end{eqnarray}
We have used modular transformation for $\vartheta$-function
in order to obtain the last expression.
The above inner product $\langle B|\cdots |B\rangle$ is computed
straightforwardly:
\begin{eqnarray}
&& \langle B|\tilde{q}^{{1\over 2}\left(L_0+\tilde{L}_0+{13\over 6}\right)}
b(2\pi i u_1)c(2\pi i Z_2)\tilde{c}(-2\pi i \bar{Z}_2)b(2\pi i u_2)
|B\rangle\nonumber\\
&=&{-i\over 8\pi}\,e^{i\pi\tilde{\tau}\over 6}
\prod_{n=1}^{\infty}(1-e^{2\pi i\tilde{\tau}n})^2\,
\bigl[g_1(u_1-Z_1|\tilde{\tau})
-g_1(u_1-Z_2|\tilde{\tau})
-g_1(u_2-Z_1|\tilde{\tau})
+g_1(u_2-Z_2|\tilde{\tau})
\bigr]\nonumber\\
&=&{-i\over 8\pi \alpha}
\,\eta(\tilde{\tau})^2\left[
{d\rho\over du}(u_1)-{d\rho\over du}(u_2)
\right],
\label{eq:BqB}
\end{eqnarray}
where we have used
\begin{eqnarray}
&&g_1(\nu|\tau):={\vartheta_1^{\prime}(\nu|\tau)\over \vartheta_1(\nu|\tau)}
=\pi \cot \pi \nu+4\pi \sum_{n=1}^{\infty}{e^{2\pi i n\tau}\over 
1-e^{2\pi i n\tau}}\sin(2\pi n \nu)\,,\\
&&L_0:=\sum_{n=1}^{\infty}n(c_{-n}b_n+b_{-n}c_n)-1\,,~~
\tilde{L}_0:=
\sum_{n=1}^{\infty}n(\tilde{c}_{-n}\tilde{b}_n+\tilde{b}_{-n}\tilde{c}_n)-1\,,
~~~~~~~~
\end{eqnarray}
and adopted the normalization as:
\begin{eqnarray}
&&|B\rangle = e^{\sum_{n=1}^{\infty}(c_{-n}\tilde{b}_{-n}
+\tilde{c}_{-n}b_{-n})}c_0^+c_1\tilde{c}_1|0\rangle\,,~~~~~
\langle 0|c_{-1}\tilde{c}_{-1}c_0\tilde{c}_0c_1\tilde{c}_1|0\rangle
=1\,.
\end{eqnarray}
We note that (\ref{eq:BqB}) corresponds to 
(5.16) in \cite{r:AKT} which was calculated 
as the correlation function at the 1-loop of {\it open} string, as expected.
We perform contour integration for $b$-ghost in (\ref{eq:Fcc})
and reduce it to evaluation of the residue at the interaction point
$\hat{z}_{\pm}$ on $u$-plane, where ${d\rho\over du}=0$,
by deforming the contour $C_1+C_2$:
\begin{eqnarray}
{\cal F}_{c\tilde{c}}&=&{2\pi\alpha_1\alpha_2\over \alpha}\,
C_{c\tilde{c}}\,\eta(\tilde{\tau})^2\,{\cal R}\,,\\
{\cal R}&=&\left(\left.{d^2\rho\over du^2}\right|_{u=\hat{z}_{\pm}}\right)^{-1}
={1\over \alpha}
\left(g^{\prime}_1(\hat{z}_{\pm}-Z_1|\tilde{\tau})
-g^{\prime}_1(\hat{z}_{\pm}-Z_2|\tilde{\tau})
\right)^{-1}.~~~~
\end{eqnarray}
In the degenerate limit, using (\ref{eq:yq}), 
the above residue ${\cal R}$ behaves as
\begin{eqnarray}
 {\cal R}&\sim&{1\over 4\alpha}\,
(\log q)^{-2}\,{q^{-1/2}\over \sin\pi\beta}\,.
\end{eqnarray}
Then, from  
$\eta(\tilde{\tau})^2=i\tilde{\tau}^{-1}\eta(-1/\tilde{\tau})^2
\sim |\tilde{\tau}|^{-1}e^{-{\pi\over 6|\tilde{\tau}|}}$
and (\ref{eq:confC}), the ghost contribution to ${\cal F}$ with 
$\phi=c\tilde{c}$ is evaluated as
\begin{eqnarray}
 {\cal F}_{c\tilde{c}}\sim 
{-\alpha_1\alpha_2\over 16\alpha^2}\,(\log q)\,
q^{13\over 12}
\left(q^{-1/2}\over \sin\pi\beta\right)^3
\end{eqnarray}
in the degenerating limit. 
On the other hand, the inner product of the right hand side of 
(\ref{eq:Bb1b2}) and $\phi=c\tilde{c}$ gives
\begin{eqnarray}
 \langle B|c_0^-c_1\tilde{c}_1|0\rangle =1\,,
\label{eq:TBgh}
\end{eqnarray}
in the ghost sector.

After combining contributions from the matter and the ghost sector,
we note that there is a constraint  $\alpha=2p^+$ in order to get physical
amplitudes \cite{KZ}. Namely, the string length parameter $\alpha$
should be identified with a light cone momentum $p^+$.
It gives extra factor $(\log q)^{-1}$ 
compared to the right hand side in (\ref{eq:B1B2m}).
(See, (5.41) in \cite{r:AKT}.)
Taking into account of it  in open string description,
${\cal F}$ (\ref{eq:calF}) is evaluated as
\begin{eqnarray}
 {\cal F}_{c\tilde{c}}&\sim&\delta_{12}\,
{-\alpha_1\alpha_2\over 16\alpha^2}\,
q^{26-c\over 24}
\left(q^{-1/2}\over \sin\pi\beta\right)^3\nonumber\\
&\sim&32\,\delta_{12}\,\alpha_1\alpha_2(\alpha_1+\alpha_2)\tau_1^{-3},
\label{eq:Ftotal}
\end{eqnarray}
where we have substituted $c=26$ as total central charge in the matter sector
 and used (\ref{eq:yq}) in the second line.

After all, using the above results: (\ref{eq:Ftotal}),(\ref{eq:TB2})
and (\ref{eq:TBgh}) for $\phi=c\tilde{c}$,
 we can evaluate the proportional constant of $B_1\star B_2\propto
\delta_{12} B_2$ for Cardy states:
\begin{eqnarray}
\left(
\langle B_1|_{\tau_1\over 2\alpha_1}b^+_0c_0^-\star  
\langle B_2|_{\tau_1\over 2\alpha_2}b^+_0c_0^-  
\right)|\phi\rangle/\langle B_2|b_0^+ c_0^- c_0^+|\phi\rangle
&\sim& 32\,\delta_{12}\,\alpha_1\alpha_2(\alpha_1+\alpha_2)
\tau_1^{-3}\,T_{B_2}^{-1}\,,
\end{eqnarray}
with regularization parameter $\tau_1$.
This implies ${\cal C}\sim 32\,\delta_{12}\,
\alpha_1\alpha_2(\alpha_1+\alpha_2)\tau_1^{-3}$
in (\ref{e_idempotency}) and is consistent with the result
in \cite{kmw2} by identifying a regularization parameter 
$\tau_1$ with $K^{-1}$ \cite{kmw2}.

\subsection{Algebra of Ishibashi states and fusion ring
\label{sec:km1}
}

Before we proceed, we point out that the idempotency relation
implies that Ishibashi state satisfies a simple algebra
with the $\star$ product.
We have discussed such relation in our previous paper \cite{km1}
by assuming the relation for the generic background.
Since it is proved in this paper, it is worth mentioning
the result again with a slight generalization.

We focus on the matter part of the
idempotency relation (\ref{e_idempotency})
which may be written as,
\begin{equation}
 |a\rangle\star |b\rangle =
q^{-{c\over 24}}\,\delta_{ab}\,T_b^{-1}|b\rangle\,.
\label{eq:m_idempotency}
\end{equation}
Here $q\,(\rightarrow 0)$ is a regularization parameter which 
was introduced in the previous subsection.
The factor $q^{-\frac{c}{24}}$ will contribute, 
when we combine it with ghost fields
with other matter sector, to a universal divergent factor.
We will therefore drop it in the following discussion 
to illuminate the nature of the algebra.

For the rational conformal field theory, the
relation between Cardy state with Ishibashi states,
(with slight generalization after \cite{Z} eq.~(2.10)),
\begin{equation}
 |a\rangle = \sum_{j}\frac{\psi_{a}^j}{\sqrt{S_{j1}}}|j\rrangle\,.
\label{eq:Cs}
\end{equation}
The coefficient $\psi_a^j$ should satisfy
the orthogonality (eqs.~(2.18) (2.19) of \cite{Z}),
\begin{equation}
 \sum_a \psi_a^i (\psi_a^j)^*=\delta_{ij}\,,\quad
 \sum_i \psi_a^i (\psi_b^i)^*=\delta_{ab}\,,\quad
\end{equation}
and also generalized Verlinde formula (2.16):
\begin{equation}
 n_{ia}{}^b=\sum_j\frac{S_{ij}}{S_{1j}}\psi^j_a (\psi^j_b)^*\,,
\end{equation}
where $n_{ia}{}^b$ are non-negative integers.
With this combination, the tension $T_a$ can be written as
\begin{equation}
 T_a=\frac{\psi^1_a}{\sqrt{S_{11}}}\,.
\end{equation}

The idempotency relation between Cardy states (in matter sector)
can be rewritten as the algebra between the Ishibashi states
$|i\rrangle$,
\begin{equation}
\label{fr}
 |i\rrangle'\star |j\rrangle' = 
\sum_k \mathcal{N}_{ij}{}^k |k\rrangle'\,,
\end{equation}
where we changed the normalization of Ishibashi states as,
\begin{equation}
 |i\rrangle' \equiv (S_{i1}S_{11})^{-1/2}|i\rrangle\,,
\end{equation}
and  ${\cal N}_{ij}^{~~k}$ is given by
\begin{eqnarray}
\mathcal{N}_{ij}^{~~k}&=&
\sum_{a}{(\psi^i_a)^*(\psi^j_a)^*\psi^k_a\over \psi^1_a}\,.
\end{eqnarray}
Eq.~(\ref{fr}) is a natural generalization of the fusion
ring for the generic BCFT, namely
the coefficient $\mathcal{N}_{ij}{}^k$ is also known to be
non-negative integers \cite{Z}.
This relation looks natural since (generalized) fusion ring
describes the number of channels in OPE of primary fields
and Ishibashi states are directly related with the 
irreducible representation.

One may summarize the observation as,
{\em Cardy states are projectors of (generalized) fusion ring}.
We believe that this nonlinear relation is a natural replacement
of Cardy condition in the first quantized language.


As a preparation of the next section, we present an application
of this result to the orbifold CFT \cite{km1}.
We consider an orbifold $\mathbf{R}^d/\Gamma$ where $\Gamma$
is a finite  group which may be nonabelian in general.
At the orbifold singularity, there exist  fractional
D-branes which are given as combinations of various
twisted sector. We apply the above idea to these fractional
D-branes.

In this setup, there is a  boundary state which belongs to
the twisted sector specified by $h\in \Gamma$,
\begin{equation}
 (X(\sigma+2\pi)-h\cdot X(\sigma))|h\rangle\!\rangle=0\,.
\end{equation}
When $\Gamma$ is nonabelian, however, such a state is not
invariant under conjugation.  Ishibashi state is, therefore,
given as a linear combination of such boundary state
which belongs to a conjugacy class $C_j$ of $\Gamma$:
\begin{equation}
 |j\rangle\!\rangle:={1\over \sqrt{r_j}}\sum_{h_j\in C_j}
|h_j\rangle\!\rangle,
\end{equation}
where 
$r_j$ is the number of elements in $C_j$.

In this case, Cardy state $|a\rangle$ is given by 
eq.~(\ref{eq:Cs}) where
the coefficients $\psi^j_a$, $S_{j1}$ are \cite{Billo}
\begin{eqnarray}
&&\psi_a^j=\sqrt{r_j\over |\Gamma|}\,\zeta^{(a)}_j,~~~~~
S_{j1}={1\over \sigma(e,h_j)},~~(h_j\in C_j)\,,
\end{eqnarray}
$\zeta^{(a)}_j$ is the character of 
an irreducible representation $a$ for $g\in C_j$,
$e$ is the identity element of $\Gamma$
and
$\sigma(e,h)$ is determined by the modular transformation of the
character $\chi_g^h(q)\equiv {\rm Tr}_{{\cal H}_g}(h\,q^{L_0-{c\over 24}})$:
\begin{eqnarray}
 \chi_e^h(q)=\sigma(e,h)\chi_h^e(\tilde{q})\,,~~~~
(q=e^{2\pi i\tau},~\tilde{q}=e^{-{2\pi i\over \tau}})\,.
\label{eq:Modular}
\end{eqnarray}
The normalization of Ishibashi state $|h\rrangle$
is specified as
$\langle\!\langle h|\tilde{q}^{{1\over
2}\left(L_0+\tilde{L}_0-{c\over 12}\right)}|h\rangle\!\rangle
= \chi_h^e(\tilde{q})\,.$
In this case, eq.~(\ref{fr}) is equivalent to
\begin{eqnarray}
&& e_i\star e_j = \sum_k{\cal N}_{ij}^{~~k}\,e_k\,,~~~~
{\cal N}_{ij}^{~~k}=\sum_{a}{r_ir_j\zeta_i^{(a)}\zeta_j^{(a)}\zeta_k^{(a)*}
\over \zeta_1^{(a)}}\,,
\label{eq:orb_alg}\\
&&e_i:=|\Gamma|\sqrt{r_i\,\sigma(e,h_i)}\,|i\rangle\!\rangle\,.
\label{eq:rescal}
\end{eqnarray}
Namely the (generalized) fusion ring is equivalent
to the group ring ${\bf C}^{[\Gamma]}$ \cite{group}.

The example in the next section is a simple example of this general
algebra.  The orbifold group $\Gamma$ is ${\bf Z}_2$ and we have
only two Ishibashi states in untwisted and twisted sector.
We write them as $|+\rrangle$ and $|-\rrangle$.
The above algebra (\ref{eq:orb_alg}) is simply,
\begin{eqnarray}
&&e_{\pm}\star e_{\pm}=e_{+}\,,\quad
e_{\pm}\star e_{\mp}=e_{-}\,,\\
&&e_{+}:=2|+\rangle\!\rangle\,,\quad
e_{-}:=2\sqrt{\sigma(e,g)}\,|-\rangle\!\rangle\,,
~~{\bf Z}_2=\{e,g\}\,,
\end{eqnarray}
(using $\sigma(e,e)=1$)
and its idempotents $P_{\pm}$ are easily obtained:
\begin{eqnarray}
P_{\pm}={1\over 2}(e_+\pm e_-)
=|+\rangle\!\rangle
\pm \sqrt{\sigma(e,g)}\,|-\rangle\!\rangle\,,
\label{eq:orb_Cardy}
\end{eqnarray}
which is the same as the Cardy states (\ref{eq:Cs})
up to overall factor.

\section{Explicit computation: toroidal and $Z_2$ orbifold compactifications}


As nontrivial examples of general arguments in the
previous section,
we calculate the $\star$ product between Ishibashi
states on torus and $Z_2$ orbifold.
We use explicit oscillator representations of three string vertices
which were formulated in \cite{HIKKO_torus}
and \cite{Itoh_Kunitomo}, respectively.
These simple examples contain nontrivial ingredients 
such as winding modes, twisted sector, cocycle factor, etc.
which make the explicit computation more interesting
compared to ${\bf R}^d$ case in \cite{kmw1}.



We use ${\bf R}^d\times T^D$
and ${\bf R}^d\times T^D/{\bf Z}_2$ as a background spacetime
and consider the HIKKO $\star$ product on them.
For the torus $T^D$, we identify
its coordinates as $X^i\sim X^i+2\pi\sqrt{\alpha'}~(i=1,\cdots,D)$
and introduce constant background
metric $G_{ij}$ and antisymmetric 
tensor $B_{ij}$.\footnote{
We mainly use the convention in \cite{KZ} although we 
introduce $\sqrt{\alpha'}$ to specify a unit length. By taking $\alpha'=1$ and 
replacing $2\pi\alpha' B_{ij}\rightarrow B_{ij}$, we recover some formulae
in \cite{KZ} for torus.}
In the case of $Z_2$ orbifold $T^D/{\bf Z}_2$,
the action of ${\bf Z}_2$ is defined by $X^i\rightarrow
-X^i~(i=1,\cdots,D)$.
The ghost sector and ${\bf R}^d$ sector of the star product are the same as 
the original HIKKO's construction \cite{HIKKO2}.
We will compute the star product of string fields of the form
$|\Phi(\alpha)\rangle=|B_D\rangle\otimes |\Phi_B(x^{\perp},\alpha)
\rangle$,
where $|B_D\rangle$ is boundary states in $D$-dimensional sector:
$T^D$ or $T^D/{\bf Z}_2$
and $|\Phi_B(x^{\perp},\alpha)\rangle=c_0^-b_0^+\,|B\rangle_{\rm
matter}\otimes|B\rangle_{\rm ghost}\otimes|\alpha\rangle$
represents a boundary state for D-brane at $x^{\perp}(\in {\bf R}^{d-p-1})$
including ghost and $\alpha$-sector.
For the ${\bf R}^d$ sector, 
conventional boundary states for D$p$-brane were proved to be
idempotent in \cite{kmw1, kmw2}:
\begin{eqnarray}
&&|\Phi_B(x^{\perp},\alpha_1)\rangle\star|\Phi_B(y^{\perp},\alpha_2)\rangle
\nonumber\\
&&~=~\delta^{d-p-1}(x^{\perp}-y^{\perp})\,\mu^2\,
{\det}^{-{d-2\over2}}(1-(\tilde{N}^{33})^2)\,
c_0^+|\Phi_B(x^{\perp},\alpha_1+\alpha_2)\rangle,~~~~
\label{eq:starRd}\\
&& \mu=e^{-\tau_0\sum_{r=1}^3\alpha_r^{-1}},~~~
\tau_0=\sum_{r=1}^3\alpha_r\log|\alpha_r|,~~~
(\alpha_3\equiv -\alpha_2-\alpha_1)\,.
\end{eqnarray}
In this section, we will focus on the matter $T^D$ or 
$T^D/{\bf Z}_2$ sector and prove a similar relation
for Cardy states on those backgrounds.

By toroidal compactification, winding mode is introduced
in addition to momentum;
the zero mode sector $|p\rangle$ changes to $|p,w\rangle$
with $p_i,w^i\in {\bf Z}$. Due to this mode,
the boundary states and the 3-string vertex should be modified.
The definition of the boundary state will be given in (\ref{eq:Ishi_T}).
The 3-string vertex should be modified to include
``cocycle factor'' such as $e^{-i\pi(p_3w_2-p_1w_1)}$.
(See, Appendix \ref{sec:starproduct} for detail.)
It is necessary to guarantee ``Jacobi identity'' with respect to 
closed string fields :
\begin{eqnarray}
(\Phi_1\star\Phi_2)\star\Phi_3
+(-1)^{|\Phi_1|(|\Phi_2|+|\Phi_3|)}(\Phi_2\star \Phi_3)\star\Phi_1
+(-1)^{|\Phi_3|(\Phi_1|+|\phi_2|)}(\Phi_3\star\Phi_1)\star\Phi_2&=&0\,,~~
\end{eqnarray}
which plays an important role to prove gauge invariance of 
the action of closed string field theory \cite{HIKKO_torus}.
It can be also derived by careful treatment of the connection condition
of light-cone type in \cite{Maeno_Takano}.
When the boundary state has non-vanishing winding number,
this cocycle factor becomes relevant.

$T^D/{\bf Z}_2$ is one of the simplest examples of orbifold
background  on which we gave a general argument
in \cite{km1} and
previous subsection (\S\ref{sec:km1}).
Cardy state $|a_{\pm} \rangle$ (\ref{eq:Cs}, \ref{eq:orb_Cardy}),
which represents fractional D-brane, is given by:
\begin{eqnarray}
 |a_{\pm} \rangle={1\over \sqrt{2}}\left(|\iota\rangle\!\rangle_u
\pm 2^{D\over 4}|\iota\rangle\!\rangle_t\right),
\end{eqnarray}
where $|\iota\rangle\!\rangle_u$ or $|\iota\rangle\!\rangle_t$ is 
a linear combination of Ishibashi states in the
untwisted or twisted sector, respectively.
The ratio of coefficients $2^{D\over 4}$
comes from the factor $\sqrt{\sigma(e,g)}$ for $T^D/{\bf Z}_2$.
We will demonstrate that string fields 
given in (\ref{eq:fractional_reg}) which are of the above form
satisfy idempotency relations 
(\ref{eq:idem1}, \ref{eq:idem2}).
It provides a consistency check for the previous general
arguments.
The oscillator computation, however, has a limitation in
determining coefficients of Ishibashi state.
They are given by determinants of
infinite rank Neumann matrices and  are divergent in general.
As a regularization, we slightly shift the interaction time 
which is specified by overlapping 
of three strings as we discussed
in \S \ref{sec:compC}.
We reduce the ratio of determinants
to the degenerating limit of the ratio
 of 1-loop amplitudes in the sense of (\ref{eq:ct_derive}).

We will also comment on  compatibility of 
idempotency relations on $T^D$ and $T^D/{\bf Z}_2$
with T-duality transformation in string field theory
which was investigated in \cite{KZ} for $T^D$.

\subsection{Star product between Ishibashi states
\label{sec:Ishi_star}
}

In this subsection, 
we first introduce Ishibashi states $|\iota\rangle\!\rangle$ 
for the backgrounds $T^D$ and  $T^D/{\bf Z}_2$
and then compute the star product between them.
In Appendix~\ref{sec:starproduct}, we give some definitions
and our convention of free oscillators.

\paragraph{Ishibashi states}
The Ishibashi states $|\iota\rangle\!\rangle$
for the torus $T^D$ are obtained by solving
$(\alpha^i_{n}+{\cal
O}^i_{~j}\tilde{\alpha}_{-n}^j)|\iota\rangle\!\rangle=0$.
${\cal O}^i_{~j}$ is an orthogonal matrix in the sense
${\cal O}^TG{\cal O}=G$.
Explicitly it is written as
\begin{eqnarray}
\label{eq:Ishi_T}
&&|\iota({\cal O},p,w)\rangle\!\rangle=
e^{-\sum_{n=1}^{\infty}{1\over n}
\alpha_{-n}^iG_{ij}{\cal O}^j_{~k}\tilde{\alpha}^k_{-n}}|p,w\rangle\,,
~~p_i=-2\pi\alpha'F_{ij}w^j\,,
\end{eqnarray}
with labels of momentum $p_i$ and winding number $w^j$ .
The antisymmetric matrix $F_{ij}$ is given by
${\cal O}=(E^T-2\pi\alpha'F)^{-1}(E+2\pi\alpha'F)$
(where $E_{ij}:=G_{ij}+2\pi\alpha'B_{ij}$).
It must be quantized in order to keep $p_i$ and $w^j$:
integers with a relation
$(1+{\cal O}^{T-1})p-(E-{\cal O}^{T-1}E^T)w=0$,
which corresponds to 
$(\alpha_0^i+{\cal O}^i_{~j}\tilde{\alpha}_0^j)|\iota\rangle\!\rangle=0$.
In particular, for Dirichlet type boundary condition,
we should set $w^i=0$ since ${\cal O}=-1$.

For $T^D/{\bf Z}_2$, there are Ishibashi states in
untwisted and twisted sectors.  For the untwisted
sector, they can be obtained by
multiplying ${\bf Z}_2$-projection to the ones for the torus: 
\begin{eqnarray}
\label{eq:Ishi_u}
 {\cal P}_u^{Z_2}|\iota({\cal O},p,w)\rangle\!\rangle_u
={1\over 2}(|\iota({\cal O},p,w)\rangle\!\rangle_u+
|\iota({\cal O},-p,-w)\rangle\!\rangle_u)\,,
\end{eqnarray}
where we add a subscript $u$ to make a distinction from 
the Ishibashi states for the torus.
For the twisted sector, we have Ishibashi states of the form:
\begin{eqnarray}
\label{eq:Ishi_tw}
|\iota({\cal O},n^f)\rangle\!\rangle_t&=&
e^{-\sum_{r=1/2}^{\infty}{1\over r}
\alpha_{-r}^iG_{ij}{\cal O}^j_{~k}\tilde{\alpha}^k_{-r}}|n^f\rangle\,.
\end{eqnarray}
The label $(n^f)^i$ takes value $0$ or $1$ and 
specifies a fixed point. This state has ${\bf Z}_2$ invariance:
${\cal P}^{Z_2}_t|\iota({\cal O},n^f)\rangle\!\rangle_t=|\iota({\cal
O},n^f)\rangle\!\rangle_t$.

\paragraph{$\star$ product}
For $T^D$ case, 
the $\star$ product of the states (\ref{eq:Ishi_T}) becomes:
\begin{eqnarray}
\label{eq:Ishi_star_T}
&&|\iota({\cal O},p_1,w_1)\rangle\!\rangle_{\alpha_1}\star 
|\iota({\cal O},p_2,w_2)\rangle\!\rangle_{\alpha_2}\nonumber\\
&&=~{\det}^{-{D\over 2}}(1-(\tilde{N}^{33})^2)
\,(-1)^{p_1 w_2}|\iota({\cal O},p_1+p_2,w_1+w_2)
\rangle\!\rangle_{\alpha_1+\alpha_2}\,,
\end{eqnarray}
where we have assigned $\alpha_r$ for each string 
(we consider the case of $\alpha_1\alpha_2>0$ here and following)
and omitted the ghost and the matter ${\bf R}^d$ sector.
Differences from ${\bf R}^d$ case \cite{kmw1} are limited to
the existence of winding mode and the cocycle factor
and the proof is similar;
we use eqs.~(\ref{eq:Ruu}) and 
(\ref{eq:Vuuu}) without ${\cal P}^{Z_2}_{u1}{\cal P}^{Z_2}_{u2}{\cal
P}^{Z_2}_{u3}$.
The cocycle factor appeared as
an extra sign factor $(-1)^{p_1 w_2}
=(-1)^{w_1 (2\pi\alpha'F) w_2}$. 
We note that this factor is irrelevant for the Dirichlet
type boundary state since we need to set $w=0$.

For $T^D/{\bf Z}_2$, we have to compute three combinations of
Ishibashi states: $({\rm untwisted})\star({\rm untwisted})$, \sloppy
$({\rm twisted})\star({\rm twisted})$ and 
$({\rm untwisted})\star({\rm twisted})$.
The first one
can be obtained by ${\bf Z}_2$-projection of the torus case
(\ref{eq:Ishi_star_T}):
\begin{eqnarray}
\label{eq:u_star_u}
&&{\cal P}^{Z_2}_u|\iota({\cal O},p_1,w_1)\rangle\!\rangle_{u,\alpha_1}\star 
{\cal P}^{Z_2}_u|\iota({\cal O},p_2,w_2)\rangle\!\rangle_{u,\alpha_2}\nonumber\\
&&={\det}^{-{D\over 2}}(1-(\tilde{N}^{33})^2)
\,{(-1)^{p_1 w_2}\over 2}\,{\cal P}^{Z_2}_u\\
&&~~~\times
\left[
|\iota({\cal O},p_1+p_2,w_1+w_2)
\rangle\!\rangle_{u,\alpha_1+\alpha_2}
+|\iota({\cal O},p_1-p_2,w_1-w_2)
\rangle\!\rangle_{u,\alpha_1+\alpha_2}
\right].\nonumber
\end{eqnarray}

The star product for 
two Ishibashi states (\ref{eq:Ishi_tw}) in the twisted sector
can be computed by the vertex operators
(\ref{eq:Rtt}, \ref{eq:Vutt}) (with appropriate permutation 
such that string $3$ is in the untwisted sector).
Using the identities among Neumann coefficients given in 
(\ref{eq:Yoneya}),
a direct computation which is similar to that in \cite{kmw1}
yields
\begin{eqnarray}
\label{eq:tt_pre}
&&|\iota({\cal O},n_1^f)\rangle\!\rangle_{t,\alpha_1}
\star |\iota({\cal O},n_2^f)\rangle\!\rangle_{t,\alpha_2}
\\
&&=~e^{{D\over 8}\tau_0(\alpha_1^{-1}+\alpha_2^{-1})}
{\det}^{-{D\over 2}}(1-(\tilde{T}^{3_u3_u})^2)
\wp{\cal P}^{Z_2}_u \sum_{p,w}
\gamma({\bf p};n_1^f,n_2^f)\,e^{\Delta E} 
\,e^{-\sum_{n>0}{1\over n}\alpha_{-n}{\cal O}\tilde{\alpha}_{-n}}
|p,w\rangle_{\alpha_1+\alpha_2}\,,\nonumber
\end{eqnarray}
where
\begin{eqnarray}
\Delta E &=& -\sum_{n>0}n^{-{1\over 2}}
(\alpha_{-n}+\tilde{\alpha}_{-n}{\cal O}^T)
\sum_{r,s=1,2}\,\sum_{m_r,l_s>0}
\tilde{T}^{3_ur}_{nm_r}
[(\tilde{T}^{\,\cdot\, 3_u}\tilde{T}^{3_u\cdot})^{-1}]^{rs}_{m_rl_s}
\tilde{T}^{s 3_u}_{l_s0}{\bf p}_+\nonumber\\
&&+{1\over 4}\left(
T^{3_u3_u}_{00}-\sum_{r,s=1,2}\,\sum_{n_r,m_s>0}
\tilde{T}^{3_ur}_{0n_r}[(1+\tilde{T})^{-1}]^{rs}_{n_rm_s}
\tilde{T}^{s 3_u}_{m_s0}
\right){\bf p}_+G^{-1}{\bf p}_+\,,\\
({\bf p}_+)_i&=&
{1\over \sqrt{2}}[(1+{\cal O}^{T-1})p-(E-{\cal O}^{T-1}E^T)w]_i\,.
\end{eqnarray}
The above peculiar exponent $\Delta E$ can be ignored because 
the coefficient of positive definite factor 
${\bf p}_+G^{-1}{\bf p}_+$ can be evaluated by
using various formulae in Appendix \ref{sec:TNeumann}  as
\begin{eqnarray}
T^{3_u3_u}_{00}-\sum_{r,s=1,2}\,\sum_{n_r,m_s>0}
\tilde{T}^{3_ur}_{0n_r}[(1+\tilde{T})^{-1}]^{rs}_{n_rm_s}
\tilde{T}^{s 3_u}_{m_s0}
&=&-\sum_{n=1}^{\infty}{2\cos^2\left({\alpha_1\over \alpha_3}n\pi\right)
\over n}\,.~~~~~~~~
\end{eqnarray}
Since it gives $-\infty$,
the ${\bf p}_+\ne 0$ terms in 
the summation in (\ref{eq:tt_pre}) is suppressed.
The constraint ${\bf p}_+=0$ implies 
$p_i=-2\pi\alpha'F_{ij}w^j$ in (\ref{eq:Ishi_T}),
 which is consistent with \sloppy
$(L_n-\tilde{L}_{-n})\left(|\iota({\cal O},n_1^f)\rangle\!\rangle_{t,\alpha_1}
\star |\iota({\cal O},n_2^f)\rangle\!\rangle_{t,\alpha_2}\right)=0
$.
This is an example of our general claim in Ref.~\cite{kmw2}
that the star product between the conformal invariant states
is again conformal invariant;
 $(L_n-\tilde{L}_{-n})|B_i\rangle=0,i=1,2,\forall n\in {\bf Z}$
~$\rightarrow$~
$(L_n-\tilde{L}_{-n})(|B_1\rangle\star |B_2\rangle)=0$.
The final form of the $\star$ product becomes,
\begin{eqnarray}
\label{eq:t_star_t}
&&|\iota({\cal O},n_1^f)\rangle\!\rangle_{t,\alpha_1}
\star |\iota({\cal O},n_2^f)\rangle\!\rangle_{t,\alpha_2}\\
&&=~e^{{D\over 8}\tau_0(\alpha_1^{-1}+\alpha_2^{-1})}
{\det}^{-{D\over 2}}(1-(\tilde{T}^{3_u3_u})^2)
\sum_{p,w,{\bf p}_+=0}(-1)^{p\,n_2^f}
\sum_m\delta^D_{n_2^f-n_1^f+w+2m,0}
|\iota({\cal O},p,w)\rangle\!\rangle_{u,\alpha_1+\alpha_2}\,.
\nonumber
\end{eqnarray}

Finally the $\star$ product between the Ishibashi states
in untwisted and twisted sectors can be computed similarly
by using the formulae in (\ref{eq:Yoneya}):
\begin{eqnarray}
&& {\cal P}_u^{Z_2}|\iota({\cal O},p_1,w_1)\rangle\!\rangle_{u,\alpha_1}
\star |\iota({\cal O},n_2^f)\rangle\!\rangle_{t,\alpha_2}
\nonumber\\
&&=~e^{{D\over 8}\tau_0(\alpha_2^{-1}-(\alpha_1+\alpha_2)^{-1})}
\,{\det}^{-{D\over 2}}(1-(\tilde{T}^{3_t3_t})^2)
\,(-1)^{p_1n_2^f}
|\iota({\cal O},[n_2^f-w_1]_{{\rm mod}\, 2})
\rangle\!\rangle_{t,\alpha_1+\alpha_2}\,.
\label{eq:u_star_t}
\end{eqnarray}
We have similar formula for [twisted(\ref{eq:Ishi_tw})] $\star$
[untwisted(\ref{eq:Ishi_u})] by appropriate replacement
in the above.

We have confirmed that Ishibashi states on
${\bf Z}_2$ orbifold (\ref{eq:Ishi_u}) and (\ref{eq:Ishi_tw})
(resp., on torus (\ref{eq:Ishi_T})) form a closed algebra with
respect to the $\star$ product as
eqs.~(\ref{eq:u_star_u}),(\ref{eq:t_star_t}) and (\ref{eq:u_star_t})
(resp., eq.~(\ref{eq:Ishi_star_T})).

\subsection{Cardy states as idempotents}


We proceed to compare the Cardy state and idempotent 
of $\star$ product algebra (fusion ring) for Ishibashi state
that we have just computed.  We note that the algebra for the
Dirichlet type boundary states are simpler since there is
no winding number and consequently the cocycle factor
in the vertex operator vanishes. Because of this simplicity
we divide our discussion into Dirichlet and Neumann type
boundary states.

\paragraph{Dirichlet type}
We start our consideration from Dirichlet type states,
namely ${\cal O}^i_{~j}=-\delta^i_j$ for the torus.
The Cardy state which describes
the Dirichlet boundary condition 
is given by a Fourier transformation of Ishibashi states
(\ref{eq:Ishi_T}) with respect to momentum $p_i$:
\begin{eqnarray}
\label{eq:B(x)}
 |B(x)\rangle&=&\left(\det(2G_{ij})\right)^{-{1\over 4}}
\sum_{p\in {\bf Z}^D}e^{-ix^ip_i}|\iota(-1,p,0)\rangle\!\rangle\,.
\end{eqnarray}
One can check that it satisfies $[\alpha^{\prime-{1\over
2}}X^i(\sigma)-x^i]_{\rm mod\, 2\pi}|B(x)\rangle=0$.
We have chosen its normalization by
\begin{eqnarray}
\langle B(x)|q^{{1\over 2}(L_0+\tilde{L}_0-{D\over 12})}|B(x')\rangle
&=&\eta(\tau)^{-D}\left[{\det}^{-{1\over 2}}(2G_{ij})
\sum_{p\in {\bf Z}^D}e^{-i(x-x')p}q^{{1\over 4}pG^{-1}p}\right]
\nonumber\\
&=&
\eta(-1/\tau)^{-D}\sum_{m\in {\bf Z}^D}
e^{-{i\over 2\pi\tau}(x-x'+2\pi m)G
(x-x'+2\pi m)}\,,
\label{eq:normD}
\end{eqnarray}
where $q=e^{2\pi i\tau}$ and 
$\eta(\tau)=q^{1\over 24}\prod_{n=1}^{\infty}(1-q^n)$.
The last representation implies that it gives
1-loop amplitude of open string whose boundaries are on D-branes at $x$
and $x'$  on the torus $T^D$.

On the other hand, from (\ref{eq:Ishi_star_T}),
the star product between them becomes
\begin{eqnarray}
&&|B(x)\rangle_{\alpha_1}\star |B(x')\rangle_{\alpha_2}\\
&&=~\delta^D([x-x^{\prime}])(2\pi)^D\left(\det(2G_{ij})\right)^{-{1\over 4}}
{\det}^{-{D\over 2}}(1-(\tilde{N}^{33})^2)\,
|B(x)\rangle_{\alpha_1+\alpha_2}\,,\nonumber
\end{eqnarray}
where
$\delta^D([x-x^{\prime}]):=\sum_{m\in{\bf Z}^D}\delta^D(x-x'+2\pi
m)=(2\pi)^{-D}\sum_{p\in{\bf Z}^D}e^{-i(x-x')p}$.
This is the idempotency relation
in \cite{kmw1} for the toroidal compactification.

For $T^D/{\bf Z}_2$, the boundary state with
${\bf Z}_2$ projection, 
${{\cal P}^{Z_2}_u}|B(x)\rangle_{u}={1\over 2}(|B(x)\rangle_{u}
+|B(-x)\rangle_{u})$ \sloppy
gives idempotents in the sense:
\begin{eqnarray}
 &&{{\cal P}^{Z_2}_u}|B(x)\rangle_{u,\alpha_1}\star {{\cal P}^{Z_2}_u}
|B(x')\rangle_{u,\alpha_2}\\
&&=\,{1\over 2}\,(\delta^D([x-x^{\prime}])+\delta^D([x+x^{\prime}]))
(2\pi)^D\left(\det(2G_{ij})\right)^{-{1\over 4}}
{\det}^{-{D\over 2}}(1-(\tilde{N}^{33})^2)\,
{{\cal P}^{Z_2}_u}|B(x)\rangle_{u,\alpha_1+\alpha_2}\,.\nonumber
\end{eqnarray}
It is clear that the combination of delta functions
is well-defined on  $T^D/{\bf Z}_2$.

At the fixed point, there are fractional D-branes.
To see them, we consider a restriction of $x$ to a fixed point $\pi n^f$,
\begin{eqnarray}
\label{eq:Bnf}
|B_{n^f}\rangle_u
&=&\left(\det(2G_{ij})\right)^{-{1\over 4}}
\sum_{p\in {\bf Z}^D}(-1)^{p\,n^f}|\iota(-1,p,0)\rangle\!\rangle_u,
\end{eqnarray}
it is ${\bf Z}_2$ invariant by itself
 ${\cal P}^{Z_2}_u|B_{n^f}\rangle_u=|B_{n^f}\rangle_u$
and is idempotent:
\begin{eqnarray}
&&|B_{n^f_1}\rangle_{u,\alpha_1}\star|B_{n^f_2}\rangle_{u,\alpha_2}\nonumber
\\
&&~=(2\pi\delta(0))^D
\delta^D_{n_1^f,n_2^f}\left(\det(2G_{ij})\right)^{-{1\over 4}}
\,{\det}^{-{D\over 2}}(1-(\tilde{N}^{33})^2)\,
|B_{n^f_1}\rangle_{u,\alpha_1+\alpha_2}.~~~~~~~
\label{eq:uu_D}
\end{eqnarray}
For the twisted sector, we can
derive from eqs.~(\ref{eq:t_star_t}),
(\ref{eq:u_star_t}) and (\ref{eq:Bnf}), 
\begin{eqnarray}
&&|B_{n_1^f}\rangle_{t,\alpha_1}\star|B_{n_2^f}\rangle_{t,\alpha_2}
\nonumber\\
&&=\delta^D_{n_1^f,n_2^f}\left(\det(2G_{ij})\right)^{1\over 4}
e^{{D\over 8}\tau_0(\alpha_1^{-1}+\alpha_2^{-1})}
{\det}^{-{D\over 2}}(1-(\tilde{T}^{3_u3_u})^2)\,
|B_{n^f_1}\rangle_{u,\alpha_1+\alpha_2}\,,\label{eq:tt_D}\\
&&|B_{n_1^f}\rangle_{u,\alpha_1}\star|B_{n_2^f}\rangle_{t,\alpha_2}
\nonumber\\
&&=\delta^D_{n_1^f,n_2^f}\left(\det(2G_{ij})\right)^{-{1\over 4}}
(2\pi\delta(0))^D\,
e^{{D\over 8}\tau_0(\alpha_2^{-1}-(\alpha_1+\alpha_2)^{-1})}
{\det}^{-{D\over 2}}(1-(\tilde{T}^{3_t3_t})^2)\,
|B_{n^f_1}\rangle_{t,\alpha_1+\alpha_2}\,,\label{eq:ut_D}~~~~~~~
\end{eqnarray}
where $|B_{n^f}\rangle_t:=
|\iota(-1,n^f)\rangle\!\rangle_t$.
These eqs.~(\ref{eq:uu_D}),(\ref{eq:tt_D}) and (\ref{eq:ut_D})
show that the Dirichlet boundary states at fixed points,
$|B_{n^f}\rangle_u$ and $|B_{n^f}\rangle_t$, form a closed algebra
with respect to the $\star$ product. 
It can be diagonalized by taking a linear combination
of the untwisted and twisted sectors:
\begin{eqnarray}
\label{eq:fractional_D}
&&|\Phi_B(n^f,x^{\perp},\alpha)\rangle_{\pm}\\
&&={1\over 2}(2\pi\delta(0))^{-D}\left(
{\det}^{1\over 4}(2G_{ij})
|B_{n^f}\rangle_u\pm c_t(2\pi\delta(0))^{D\over 2}
|B_{n^f}\rangle_t\right)\otimes|\Phi_B(x^{\perp},\alpha)\rangle,
\nonumber
\end{eqnarray}
where we have included a string field
$|\Phi_B(x^{\perp},\alpha)\rangle$,
which is a contribution from the other part of
matter sector ${\bf R}^d$ and ghost sector. It
is essentially a  boundary state for D$p$-brane.
The coefficient of the boundary states in the twisted sector
is given by a ratio of the determinants of Neumann matrices:
\begin{eqnarray}
 c_t&:=&\sqrt{{\cal C}\over {\cal C}'}
=\left(
e^{-{\tau_0\over 4}(\alpha_1^{-1}+\alpha_2^{-1})}\,{
\det(1-(\tilde{T}^{3_u 3_u})^2)
\over \det(1-(\tilde{N}^{3 3})^2)}
\right)^{D\over 4},
\label{eq:ct}
\\
{\cal C}&=&\mu^2\,{\det}^{-{d+D-2\over 2}}(1-(\tilde{N}^{33})^2)\,,~~~~
\mu=e^{-\tau_0(\alpha_1^{-1}+\alpha_2^{-1}-(\alpha_1+\alpha_2)^{-1})},
~~~\\
 {\cal C}'&=&\mu^2\,e^{{D\over 8}\tau_0(\alpha_1^{-1}+\alpha_2^{-1})}
{\det}^{-{D\over 2}}(1-(\tilde{T}^{3_u 3_u})^2)\,
{\det}^{-{d-2\over 2}}(1-(\tilde{N}^{33})^2)\,.
\end{eqnarray}
They satisfy idempotency relations of the following form:
\begin{eqnarray}
 &&|\Phi_B(n_1^f,x^{\perp},\alpha_1)\rangle_{\pm}
\star |\Phi_B(n_2^f,y^{\perp},\alpha_2)\rangle_{\pm}\nonumber\\
&&~~~=~\delta^D_{n_1^f,\,n_2^f}\,\delta^{d-p-1}(x^{\perp}-y^{\perp})\,
{\cal C}\,c_0^+|\Phi_B(n_1^f,x^{\perp},\alpha_1+\alpha_2)\rangle_{\pm}\,,
\label{eq:idem1}
\\
\label{eq:idem2}
 &&|\Phi_B(n_1^f,x^{\perp},\alpha_1)\rangle_{\pm}
\star |\Phi_B(n_2^f,y^{\perp},\alpha_2)\rangle_{\mp}=0\,,
\end{eqnarray}
where ${\cal C}$ was computed in \cite{kmw2}
and is proportional to
$K^3\alpha_1\alpha_2(\alpha_1+\alpha_2)$
for $d+D=26$ with cutoff parameter $K$.
In the above computation, we have used the relation of determinants
of Neumann matrices:
\begin{eqnarray}
\label{eq:CG_T}
{\det}^{-{1\over 2}}(1-(\tilde{N}^{33})^2)
=e^{{1\over 8}\tau_0(\alpha_2^{-1}-(\alpha_1+\alpha_2)^{-1})}
{\det}^{-{1\over 2}}(1-(\tilde{T}^{3_t 3_t})^2)\,,
\end{eqnarray}
which can be proved analytically 
by using Cremmer-Gervais identity
as in Ref.~\cite{kmw2}.   Outline of the proof is given in 
Appendix \ref{sec:CG}.
It can be also checked numerically
by truncating the size of Neumann
matrices to $L\times L$. ($L\sim K$)

As for the coefficient $c_t$ (\ref{eq:ct})
in front of the twisted sector,
it can be evaluated by another regularization
as \S \ref{sec:compC} (See, Appendix \ref{sec:1-loop} for detail.)
The result is given in (\ref{eq:ct_derive}):
\begin{eqnarray}
\label{eq:ct_value}
 c_t(2\pi\delta(0))^{D\over 2}
=2^{D\over 4}\left(\det(2G)\right)^{1\over 4}
=\sqrt{\sigma(e,g)}\left(\det(2G)\right)^{1\over 4}\,,
\end{eqnarray}
where $\sigma(e,g)=2^{D\over 2}$ 
is the Modular transformation matrix defined in (\ref{eq:Modular})
and is given in \cite{Billo}.
This implies that the idempotents (\ref{eq:fractional_D})
is proportional to the Cardy state for the fractional D-branes,
\begin{eqnarray}
\label{eq:fractional_reg}
|\Phi_B(n^f,x^{\perp},\alpha)\rangle_{\pm}
={1\over 2}\,{{\det}^{1\over 4}(2G)\over (2\pi\delta(0))^D}
\left(|B_{n^f}\rangle_u\pm 2^{D\over 4}
|B_{n^f}\rangle_t\right)\otimes|\Phi_B(x^{\perp},\alpha)\rangle,
~~~~~~
\end{eqnarray}
after a proper regularization.

\paragraph{Neumann type}
We call the boundary states with
${\cal O}^i_{~j}\ne -\delta^i_j$ as Neumann type
while they may have mixed boundary condition in general.
As we wrote, the derivation of idempotent for such states
is slightly more nontrivial because of the cocycle factor
in the vertex.

We start again from the toroidal compactification and
consider a particular linear combination of
Ishibashi states (\ref{eq:Ishi_T}) of the form:
\begin{eqnarray}
\label{eq:Bq}
 |B(q),F\rangle:={\det}^{-{1\over 4}}(2G_O^{-1})
\sum_{w}e^{-iqw+i\pi wF_uw}
|\iota({\cal O},-2\pi\alpha'Fw,w)\rangle\!\rangle\,,
\end{eqnarray}
where we denote $(F_u)_{ij}= 2\pi\alpha'F_{ij}$ for $i<j$ and 
$(F_u)_{ij}=0$ for $i\ge j$.
As we explained, $2 \pi \alpha' F_{ij}$ should be quantized 
for the consistency with the momentum quantization.
We have chosen the normalization factor ${\det}^{-{1\over 4}}(2G_O^{-1})$ by
\begin{eqnarray}
\langle B(q'),F|\,e^{\pi i\tau\left(L_0+\tilde{L}_0-{D\over 12}\right)}
|B(q),F\rangle
&=&{\det}^{-{1\over 2}}(2G_O^{-1})\,\eta(\tau)^{-D}\sum_{w^i}
e^{-i(q-q')_iw^i}e^{{\pi i\tau\over 2}w^iG_{Oij}w^j}\nonumber\\
&=&\eta(-1/\tau)^{-D}\sum_{m}
e^{-{i\over 2\pi \tau}(q-q'+2\pi m)_iG_O^{ij}(q-q'+2\pi m)_j}\,,
\label{eq:normN}
\end{eqnarray}
as (\ref{eq:normD}).
Here $q_i\equiv q_i+2\pi$ corresponds 
to Wilson line on the D-brane and
${G_O}_{ij}:=G_{ij}-(E+2\pi\alpha' F)_{ik}G^{kl}(E^T-2\pi\alpha' F)_{lj}
$ is the open string metric.

We compute the $\star$ product of
(\ref{eq:Bq}) using (\ref{eq:Ishi_star_T}):
\begin{eqnarray}
&&|B(q_1),F\rangle_{\alpha_1}\star|B(q_2),F\rangle_{\alpha_2}
\\
&&~=\delta^D([q_1-q_2])\,
(2\pi)^D{\det}^{-{1\over 4}}(2G_O^{-1})
\,{\det}^{-{D\over 2}}(1-(\tilde{N}^{33})^2)\,
|B(q_1),F\rangle_{\alpha_1+\alpha_2},\nonumber
\end{eqnarray}
which is the idempotency relation for $T^D$.
We note that due
the phase factor $e^{i\pi wF_u w}$ in (\ref{eq:Bq}),
Cardy state is not the Fourier transform of the Ishibashi state.
It is necessary to cancel the cocycle
factor in the 3-string vertex (\ref{eq:Vuuu}).
It is also necessary to keep
T-duality symmetry in closed string field theory on 
the torus $T^D$, (see, eq.~(\ref{eq:upsilon}) in particular).

For the orbifold,
we can check that
${\cal P}^{Z_2}_u|B(q),F\rangle_{u}={1\over 2}(
|B(q),F\rangle_{u}+|B(-q),F\rangle_{u})$
is idempotent in the untwisted sector
on  $T^D/{\bf Z}_2$:
\begin{eqnarray}
 &&{{\cal P}^{Z_2}_u}|B(q),F\rangle_{u,\alpha_1}\star {{\cal P}^{Z_2}_u}
|B(q'),F\rangle_{u,\alpha_2}\\
&&~~=~{1\over 2}(\delta^D([q-q^{\prime}])+\delta^D([q+q^{\prime}]))
(2\pi)^D{\det}^{-{1\over 4}}(2G_O^{-1})
{\det}^{-{D\over 2}}(1-(\tilde{N}^{33})^2)\,
{{\cal P}^{Z_2}_u}|B(q),F\rangle_{u,\alpha_1+\alpha_2}\,.\nonumber
\end{eqnarray}

Mixing with the twisted sector occurs when
the Wilson line takes special values,
$q_i=\pi m^f_i$ ($m_i^f=0,1$) for the untwisted sector:
\begin{eqnarray}
\label{eq:Bmf}
&&|B_{m^f},F\rangle_u
={\det}^{-{1\over 4}}(2G_O^{-1})
\sum_{w}(-1)^{m^f w+wF_uw}
|\iota({\cal O},-2\pi\alpha'Fw,w)\rangle\!\rangle_u\,.~~~~~~~~
\end{eqnarray}
These states are by themselves ${\bf Z}_2$ invariant: 
${\cal P}^{Z_2}_u|B_{m^f},F\rangle_u=|B_{m^f},F\rangle_u$.
The star product between them is,
\begin{eqnarray}
&&|B_{m^f_1},F\rangle_{u,\alpha_1}\star |B_{m^f_2},F\rangle_{u,\alpha_2}
\label{eq:uu_N}\\
&&~=\delta^D_{m^f_1,m^f_2}{\det}^{-{1\over 4}}(2G_O^{-1})(2\pi\delta(0))^D
{\det}^{-{D\over 2}}(1-(\tilde{N}^{33})^2)\,
|B_{m^f_1},F\rangle_{u,\alpha_1+\alpha_2}\,.\nonumber
\end{eqnarray}
In the twisted sector, we consider a particular
linear combination  of Ishibashi states (\ref{eq:Ishi_tw})
such as
\begin{eqnarray}
 \label{eq:Bmft}
 |B_{m^f},F\rangle_t:=2^{-{D\over 2}}\sum_{n^f_i=0,1}
(-1)^{m^fn^f+n^fF_un^f}
|\iota({\cal O},n^f)\rangle\!\rangle_t\,,
\end{eqnarray}
which is a generalization of the twisted Neumann boundary state 
in \cite{oshi}.
Here, we have also multiplied the phase factor $(-1)^{n^fF_un^f}$
as in the untwisted sector (\ref{eq:Bmf})
for the idempotency.
We can derive the $\star$ product formulae
\begin{eqnarray}
&&|B_{m_1^f},F\rangle_{t,\alpha_1}
\star |B_{m^f_2},F\rangle_{t,\alpha_2}\\
&&=\delta^D_{m^f_1,\,m^f_2}
{\det}^{1\over 4}(2G_O^{-1})\,
e^{{D\over 8}\tau_0(\alpha_1^{-1}+\alpha_2^{-1})}
{\det}^{-{D\over 2}}(1-(\tilde{T}^{3_u3_u})^2)
|B_{m^f_1},F\rangle_{u,\alpha_1+\alpha_2},\nonumber\\
&&|B_{m_1^f},F\rangle_{u,\alpha_1}
\star |B_{m_2^f},F\rangle_{t,\alpha_2}\\
&&=\delta_{m_1^f,\,m_2^f}{\det}^{-{1\over 4}}(2G_O^{-1})
(2\pi\delta(0))^D
e^{{D\over 8}\tau_0(\alpha_2^{-1}-(\alpha_1+\alpha_2)^{-1})}
{\det}^{-{D\over 2}}(1-(\tilde{T}^{3_t3_t})^2)\,
|B_{m_1^f},F\rangle_{t,\alpha_1+\alpha_2}
,\nonumber
\end{eqnarray}
{}from eqs.~(\ref{eq:t_star_t}),(\ref{eq:u_star_t}),(\ref{eq:Bmf})
and (\ref{eq:Bmft}).
Using the above formulae, noting eq.~(\ref{eq:CG_T}),
we obtain idempotents which include the twisted sector:
\begin{eqnarray}
\label{eq:fractional_DN}
&&|\Phi_B(m^f,F,x^{\perp},\alpha)\rangle_{\pm}
={1\over 2}\,{{\det}^{1\over 4}(2G_O^{-1})\over (2\pi\delta(0))^D}
\left(|B_{m^f},F\rangle_u\pm 2^{D\over 4}\,
|B_{m^f},F\rangle_t\right)\otimes|\Phi_B(x^{\perp},\alpha)\rangle\,.~~~
\end{eqnarray}
Here we again include the extra matter fields on
${\bf R}^d$ and ghost sector:
 $|\Phi_B(x^{\perp},\alpha)\rangle$.  We evaluate
the ratio of determinants $c_t$ (\ref{eq:ct})
using the regularization given by (\ref{eq:BBuN}) instead of 
(\ref{eq:BBu}) because we are treating Neumann type boundary
states.
Their star product becomes idempotent as expected,
\begin{eqnarray}
 &&|\Phi_B(m_1^f,F,x^{\perp},\alpha_1)\rangle_{\pm}
\star |\Phi_B(m_2^f,F,y^{\perp},\alpha_2)\rangle_{\pm}\nonumber\\
&&~~~=~\delta^D_{m_1^f,\,m_2^f}\,\delta^{d-p-1}(x^{\perp}-y^{\perp})\,
{\cal C}\,c_0^+|\Phi_B(m_1^f,x^{\perp},\alpha_1+\alpha_2)\rangle_{\pm}\,,
\label{eq:idem1N}
\\
\label{eq:idem2N}
 &&|\Phi_B(m_1^f,x^{\perp},\alpha_1)\rangle_{\pm}
\star |\Phi_B(m_2^f,y^{\perp},\alpha_2)\rangle_{\mp}=0\,.
\end{eqnarray}

\subsection{Comments on T-duality}


We have seen that the Dirichlet type idempotent 
and the Neumann type one are constructed in slightly different
manner due to the cocycle factor.
They are related, however, by T-duality transformation
and we would like to see explicitly how the difference
can be absorbed.  In this subsection we follow
the argument of \cite{KZ}.

A key ingredient is the existence of the following operator
$\mathcal{U}_g^\dagger$,
\begin{eqnarray}
\label{eq:T-dualU}
 {\cal U}_g^{\dagger}|A\star B\rangle_E
&=&|({\cal U}_g^{\dagger}A)\star({\cal U}_g^{\dagger}B)\rangle_{g(E)}\,.
\end{eqnarray}
Here the subscripts of the ket: $E$ and $g(E)$ specify the constant
background $E=G+2\pi \alpha' B$ and its T-duality transformation
specified by $g\in O(D,D;{\bf Z})$:
\begin{eqnarray}
&&g(E):=(aE+b)(cE+d)^{-1},\\
&&g=\left(
\begin{array}[tb]{cc}
 a & b \\
 c & d
\end{array}
\right),
~~~~g^T J g=g\,,~J:=\left(
\begin{array}[tb]{cc}
 0&1_D \\
1_D&0
\end{array}
\right).
\end{eqnarray}
The operator ${\cal U}_g$ is defined by \cite{KZ}:
\begin{eqnarray}
&& {\cal U}_g=U_g\,\Upsilon(g,\hat{p},\hat{w})\,,~~
U_g^{\dagger}|p,w\rangle_E=|ap+bw,cp+dw\rangle_{g(E)},\\
&&U_g^{\dagger}\alpha_n(E) U_g=(d-cE^T)^{-1}\alpha_n(g(E))\,,\\
&&U_g^{\dagger}\tilde{\alpha}_n(E) U_g=(d+cE)^{-1}
\tilde{\alpha}_n(g(E))\,,~~~~~\\
&&\Upsilon(g,\hat{p},\hat{w})=\exp(i\pi(\hat{p}(dc^T)_u\hat{p}
+\hat{w}(ba^T)_u\hat{w}+\hat{w}bc^T\hat{p}))\,,
\label{eq:upsilon}
\end{eqnarray}
where matrices in the exponent of $\Upsilon$ with subscript $u$ are
defined by $(A_u)_{ij}= A_{ij}\,(i<j)$ and $(A_u)_{ij}=0\,(i\ge j)$.
In particular, we consider a class of  $O(D,D;{\bf Z})$-transformation
of the form:
\begin{eqnarray}
\label{eq:gDN}
 g_{\rm DN}&=&\left(
\begin{array}[tb]{cc}
 -2\pi\alpha'F& 1\\
1&0
\end{array}
\right)\,,~~~~~~(2\pi\alpha'F_{ij}=-2\pi\alpha'F_{ji}\in\mathbf{Z}).
\end{eqnarray}
They give T-duality transformations between the idempotents 
\begin{eqnarray}
{\cal U}_{g_{\rm DN}}^{\dagger}|B(x^i)\rangle_{E}=
|B(q_i=x^i),F\rangle_{g_{\rm DN}(E)}\,,
\end{eqnarray}
on the torus. 
Note that the original metric $G$ is mapped to the inverse open string metric 
$G^{-1}_O$ by the transformation $g_{\rm DN}$:
\begin{eqnarray}
G=(E^{\prime T}-2\pi\alpha'F)^{-1}G'(E'+2\pi\alpha' F)^{-1}=G^{\prime-1}_O
\,,~~~~~E'=g_{\rm DN}(E)\,.
\end{eqnarray}
Indeed, this is consistent with general property 
of the $\star$ product (\ref{eq:T-dualU}).

We can extend such an analysis to 
$T^D/{\bf Z}_2$ case. 
We define a unitary operator ${\cal U}_{g_{\rm DN}}$
which represent the action of
 $g_{\rm DN}$ (\ref{eq:gDN}) to the twisted sector:
\begin{eqnarray}
{\cal U}_{g_{\rm DN}}^{\dagger}\alpha_r(E) {\cal U}_{g_{\rm DN}}&=&-E^{T-1}
\alpha_r(g_{\rm DN}(E)),\\
{\cal U}_{g_{\rm DN}}^{\dagger}\tilde{\alpha}_r(E) {\cal U}_{g_{\rm DN}}
&=&E^{-1}\tilde{\alpha}_r(g_{\rm DN}(E)),
\end{eqnarray}
where $\alpha_r(E),~(r\in {\bf Z}+{1/2})$
 is the oscillator on the background $E$.
For the oscillator vacuum $|n^f\rangle_E$, we
define 
\begin{eqnarray}
 {\cal U}^{\dagger}_{g_{\rm DN}}|n^f\rangle_{E}
&=&2^{-{D\over 2}}\sum_{m^f_i=0,1}(-1)^{n^f_im^f_i+m^f_i(F_u)_{ij}m^f_j}
|n^f\rangle_{g_{\rm DN}(E)}\,.
\end{eqnarray}
Then, with ${\bf Z}_2$ projection, we can prove
\begin{eqnarray}
 {\cal U}^{\dagger}_{g_{\rm DN}}|A\star B\rangle_E
=|( {\cal U}^{\dagger}_{g_{\rm DN}}A)
\star({\cal U}^{\dagger}_{g_{\rm DN}}B)\rangle_{g_{\rm DN}(E)},
\end{eqnarray}
not only in the untwisted sector
but also in the twisted sector
by investigating reflectors (\ref{eq:Ruu}),(\ref{eq:Rtt}) 
and 3-string vertices (\ref{eq:Vuuu}),(\ref{eq:Vutt}).
This implies that we obtain Neumann type idempotents 
(\ref{eq:fractional_DN}) from Dirichlet type 
(\ref{eq:fractional_D}) by $ {\cal U}^{\dagger}_{g_{\rm DN}}$:
\begin{eqnarray}
 {\cal U}^{\dagger}_{g_{\rm DN}}|\Phi_B(n^f_i,x^{\perp},\alpha)\rangle_{\pm,E}
=|\Phi_B(m^f_i=n^f_i,F,x^{\perp},\alpha)\rangle_{\pm,\,g_{\rm DN}(E)}\,.
\end{eqnarray}

\section{Deformation of the algebra by $B$ field}



In this section, we consider a  deformation
along the transverse directions by the introduction of $B$ field.
In Seiberg-Witten limit, it induces noncommutativity
to the ring of functions on these directions.
Since our equation, $\Phi\star \Phi=\Phi$ formally
resembles GMS soliton equation, it is curious how our
star product is modified in such limit.

In particular, the algebra of Ishibashi state in transverse dimension
was,
\begin{equation}
 |p\rrangle\star | q\rrangle = a( p,
q)|{p+q}\rrangle\,,\quad
(a(p,q)=1)
\end{equation}
when there is no $B$ field.  In order to obtain a projector 
for this algebra, we perform a Fourier transformation
$|x^\perp\rangle=\int dk \,e^{ikx^\perp}|k\rrangle$,
which combines Ishibashi states to Cardy state, 
and this is identical to the the boundary state 
for the transverse direction.

A naive guess is that the product becomes Moyal product, namely
$a(p,q)$ becomes $\exp\left(-\frac{i}{2} p_i\theta^{ij} q_j\right)$.
This can not, however, be the case since the closed string star
product is commutative. We will see that in a {\em specific} setup
which we are going to consider,
the factor becomes
\begin{equation}
 a(p,q)=\frac{\sin(-\beta\lambda)}{-\beta\lambda}
\frac{\sin((1+\beta)\lambda)}{(1+\beta)\lambda}\,,
\quad \lambda=-{1\over 2}\,p_i\theta^{ij}q_j\,,
\end{equation}
for HIKKO type star product in the Seiberg-Witten limit.  
If we expand in terms of $\lambda$, it is easy to see that
this expression reduces to $1$ when $\theta\rightarrow 0$.
It is commutative and non-associative which are
the basic properties of closed string star product.

If we  know the boundary state in the presence of $B$ field
in the transverse dimensions,  our computation would be
straightforward since the definition of the star product itself
remains the same.  Actually, however, the boundary state
which corresponds to GMS soliton is not known.
Namely, the treatment of the massive particles is
difficult.  Such modes can be decoupled from zero-mode 
only when Seiberg-Witten limit is taken.

Therefore, we are going to take the following path
to obtain the deformation of the algebra,
\begin{enumerate}
 \item define an operator $V_\theta$ (\ref{eq:KTop})
       which describes the deformation by $B$ 
       field and apply that operator to Ishibashi states,
       $|p\rrangle'=V_\theta |p\rrangle$,
 \item calculate $\star$ product between these states
       $|p\rrangle'\star |q\rrangle'$,
 \item and take Seiberg-Witten limit.
\end{enumerate}
Actually the state obtained in the step 1 does not satisfy
the conformal invariance $(L_n-\tilde L_{-n})|B\rangle=0$.
It means that they are not, precisely speaking, the boundary
states.  Instead, we will see that the deformed Ishibashi state
is equivalent to Neumann type boundary state with tachyon
vertex insertion (\ref{eq:VD=VBF}).
It may imply that our computation in the following
should be related to the loop correction factor
in noncommutative Yang-Mills theory.

\subsection{A deformation of boundary state in the presence of $B$ field}

Let us first introduce ``KT-operator'' $V_{\theta}=e^M$
\cite{Kawano_Takahashi, KT2},
which defines the deformation associated with the  noncommutativity 
for the constant $B$-field background in Witten's open string
field theory and 
HIKKO open-closed string
field theory. In that context, it was demonstrated that
this operator $V_{\theta}$ transforms
open string fields on $B=0$ background to that on $B\ne 0$.
The KT operator $V_{\theta}$ on a constant metric $g_{ij}$ background
is given by
\begin{eqnarray}
\label{eq:KTop}
 V_{\theta}&=&\exp\left(-{i\over 4}\oint d\sigma\oint d\sigma'
P_i(\sigma)\theta^{ij}
\epsilon(\sigma-\sigma')
P_j(\sigma')\right)
\end{eqnarray}
where $P_i(\sigma)={1\over 2\pi}\left[
\hat{p}_i+{1\over\sqrt{2\alpha'}}\sum_{n\ne 0,\,n\in {\bf Z}}g_{ij}
\left(\alpha_n^j
e^{in\sigma}+\tilde{\alpha}_n^je^{-in\sigma}\right)
\right]$, 
and $\epsilon(x)$ is the sign function. Formally, we get
\begin{equation}
V_{\theta}\partial_{\sigma}X^i(\sigma)
V^{-1}_{\theta}=\partial_{\sigma}X^i(\sigma)-\theta^{ij}P_j(\sigma) 
\end{equation}
by canonical commutation relation, and therefore, we can expect that
the operator $V_{\theta}$ induces a map from Dirichlet boundary state
to Neumann one with constant flux.\footnote{
This operator was also obtained using path integral
formulation \cite{Oku}
in the process of constructing boundary state
for D$p$-brane from that for D-instanton.
}

A subtlety in (\ref{eq:KTop}) is
how to define
$\epsilon(\sigma-\sigma')$ or $\oint
d\sigma\oint d\sigma'$
since we need to impose
the periodicity of {\it closed} strings 
$P_i(\sigma+2\pi)=P_i(\sigma)$.
Here, we introduce a cut $\sigma_c$ and
set the integration region to $\sigma\in [\sigma_c,2\pi+\sigma_c]$.
Then, by taking normal ordering using a formula given in (\ref{eq:KP}),
an explicit oscillator representation of 
KT operator $V_{\theta}$ (\ref{eq:KTop})
becomes,
\begin{eqnarray}
V_{\theta,\sigma_c}&:=&\exp\left(-{i\over 4}
\int_{\sigma_c}^{2\pi+\sigma_c} d\sigma
\int_{\sigma_c}^{2\pi+\sigma_c} d\sigma'
P_i(\sigma)\theta^{ij}
\epsilon(\sigma-\sigma')
P_j(\sigma')\right)\nonumber\\
&=&\left(\det\left(1-C\right)\right)^{-{1\over 2}}
e^{{1\over 2}DN(1-C)^{-1}D^T}
e^{-{1\over 2}a^{\dagger}NC(1+C)^{-1}a^{\dagger}+D(1+C)^{-1}a^{\dagger}}
\nonumber\\
&&\times\,
e^{-a^{\dagger}\log(1-C)a}
e^{{1\over 2}aNC(1-C)^{-1}a+DN(1-C)^{-1}a},\label{eq:KTop2}
\end{eqnarray}
where 
\begin{eqnarray}
&&a=\left(
\begin{array}[tb]{c}
 {(e\alpha_n)^a\over \sqrt{n}}\\
 {(e\tilde{\alpha}_n)^a\over \sqrt{n}}
\end{array}
\right),~
a^{\dagger}=\left(
\begin{array}[tb]{c}
 {(e\alpha_{-n})^a\over \sqrt{n}}\\
 {(e\tilde{\alpha}_{-n})^a\over \sqrt{n}}
\end{array}
\right),~(n\ge 1);
~~
g_{ij}=e_i^a\eta_{ab}e_j^b\,,
~g^{ij}=\tilde{e}^i_a\eta^{ab}\tilde{e}^j_b\,,
~e_i^a\tilde{e}_a^j=\delta_i^j,~~~~~\\
&&C=-C^T=-{1\over 4\pi\alpha'}\left(
\begin{array}[tb]{cc}
 (e\theta e)^{ab}\delta_{n,m}& 0\\
0&-(e\theta e)^{ab}\delta_{n,m}
\end{array}
\right),\\
&&N=N^T=\left(
\begin{array}[tb]{cc}
 0& \eta_{ab}\delta_{n,m}\\
\eta_{ab}\delta_{n,m}&0
\end{array}
\right),
~~~~
D=-{1\over 2\pi\sqrt{2\alpha'}}\,\hat{p}_i\,\theta^{ij}\left(e_j^a\,
{e^{-im\sigma_c}\over \sqrt{m}},
-e_j^a\,{e^{im\sigma_c}\over \sqrt{m}}\right).
\end{eqnarray}
By multiplying $V_{\theta,\sigma_c}$ (\ref{eq:KTop2}) to
the  Dirichlet type Ishibashi state with momentum $p$:
$|p\rangle\!\rangle_D:=e^{\sum_{n\ge 1}{1\over n}\alpha^i_{-n}
g_{ij}\tilde{\alpha}^j_{-n}}|p\rangle$, we obtain
\begin{eqnarray}
&&V_{\theta,\sigma_c}|p\rangle\!\rangle_D=
\left[\det\left(1-(2\pi\alpha')^{-1}g\theta\right)\right]^{-\sum_{n\ge 1}1}\,
e^{-\alpha'pG^{-1}_{\theta}p\,\sum_{n\ge 1}{1\over n}}\\
&&~~~~~~~~~~~~~~~~
\times\exp\left(-\sum_{n=1}^{\infty}{1\over n}
\alpha_{-n}g{\cal O}_{\theta}\tilde{\alpha}_{-n}
+\sum_{n=1}^{\infty}(\lambda_n\alpha_{-n}+\tilde{\lambda}_n
\tilde{\alpha}_{-n})\right)|p\rangle,\nonumber\\
&&{\cal O}_{\theta}=(g+2\pi\alpha'\theta^{-1})^{-1}
(g-2\pi\alpha'\theta^{-1})\,,~~~
G^{-1}_{\theta}=(g-2\pi\alpha'\theta^{-1})^{-1}
g(g+2\pi\alpha'\theta^{-1})^{-1},\\
&&(\lambda_m,\tilde{\lambda}_m)
=\sqrt{2\alpha'}\,p
\left(
(g-2\pi\alpha'\theta^{-1})^{-1}g\,{e^{-im\sigma_c}\over m}
,(g+2\pi\alpha'\theta^{-1})^{-1}g\,{e^{im\sigma_c}\over m}
\right).
\end{eqnarray}
We redefine the normalization of this state as
\begin{eqnarray}
\label{eq:KT_D}
 \hat{V}_{\theta,\sigma_c}|p\rangle\!\rangle_D:=
\left[\det\left(1-(2\pi\alpha')^{-1}g\theta\right)\right]^{\sum_{n\ge 1}1}
e^{\alpha'pG^{-1}_{\theta}p\,\sum_{n\ge 1}{1\over n}}\,
V_{\theta,\sigma_c}|p\rangle\!\rangle_D,
\end{eqnarray}
so that $\langle p'|\hat{V}_{\theta,\sigma_c}|p\rangle\!\rangle_D=
(2\pi)^d\delta^d(p'-p)$. Then, we find an identity
\begin{eqnarray}
\label{eq:VD=VBF}
\hat{V}_{\theta,\sigma_c}|p\rangle\!\rangle_D
&=& V_p(\sigma_c)|B(F_{ij}=-(\theta^{-1})_{ij})\rangle\,,
\end{eqnarray}
where
\begin{eqnarray}
 |B(F)\rangle=e^{-
\sum_{n=1}^{\infty}{1\over n}\alpha_{-n}g{\cal O}\tilde{\alpha}_{-n}}
|p=0\rangle
\end{eqnarray}
with ${\cal O}=(g-2\pi\alpha'F)^{-1}(g+2\pi\alpha'F)$,
is the Neumann boundary state with constant flux $F$ and 
\begin{eqnarray}
&&V_k(\sigma)={\cal N}_k:e^{ikX(\sigma)}:\nonumber\\
&&=
e^{{1\over 2}\alpha'kG^{-1}_Ok\sum_{n\ge 1}{1\over n}}\,
e^{k\sum_{n=1}^{\infty}\left(\alpha'\over 2\right)^{1/2}{1\over n}
\left(
\alpha_{-n}e^{-in\sigma}
+\tilde{\alpha}_{-n}e^{in\sigma}
\right)}
e^{ik\hat{x}}\,
e^{-k\sum_{n=1}^{\infty}\left(\alpha'\over 2\right)^{1/2}{1\over n}
\left(\alpha_ne^{in\sigma}+\tilde{\alpha}_ne^{-in\sigma}\right)
},~~~~~~~
\label{eq:tach}
\end{eqnarray}
where $G_O^{-1}=(g+2\pi\alpha'F)^{-1}
g(g-2\pi\alpha'F)^{-1}$ is the open string metric,
represents the tachyon vertex operator at $\sigma$
with momentum $k_i$.

The above identity (\ref{eq:VD=VBF}) implies that 
the KT operator (\ref{eq:KTop2}) maps 
the Dirichlet type Ishibashi state of momentum $p$ to 
Neumann boundary states {\it with} tachyon vertex with momentum $p$,
where the position of the tachyon insertion $\sigma_c$ corresponds to
the cut 
in the definition of the exponent of (\ref{eq:KTop2}).
This combination was investigated as a fluctuation 
around 
boundary states 
in \cite{kmw1,kmw2} and can be used to
calculate their star product in the following.

\subsection{$\star$ product of deformed Ishibashi state}
Let us proceed to the step 2, namely the computation of
 the $\star$ product of
$\hat{V}_{\theta,\sigma_c}|p\rangle\!\rangle_D$ (\ref{eq:KT_D}).
We use eqs.~(4.6) and (4.7) in \cite{kmw1} to give
\begin{eqnarray}
 &&\hat{V}_{\theta,\sigma_c}|p_1\rangle\!\rangle_{D,\alpha_1}
\star\hat{V}_{\theta,\sigma_c}|p_2\rangle\!\rangle_{D,\alpha_2}\nonumber\\
&&={\cal N}_{12}\,{\det}^{-{d\over 2}}(1-(\tilde{N}^{33})^2)
\oint {d\sigma_1\over 2\pi}
\oint {d\sigma_2\over 2\pi}|e^{i\sigma^{(1)}(\sigma_1)}
-e^{i\sigma^{(2)}(\sigma_2)}|^{2\alpha'p_1G^{-1}_Op_2}\,
e^{i\Theta_{12}}\nonumber\\
&&~~~~\times\,\wp\, e^{\sum_{n\ge 1}(\lambda_n^{(12)}\alpha_{-n}
+\tilde{\lambda}_n^{(12)}\tilde{\alpha}_{-n})-{\sum_{n\ge 1}{1\over n}
\alpha_{-n}G{\cal O}\tilde{\alpha}_{-n}}}|p_1+p_2\rangle_{\alpha_1+\alpha_2}\,,
\label{eq:VDstar}
\end{eqnarray}
where we have assigned $\alpha_1,\alpha_2$ 
($\alpha_1\alpha_2>0$)
and omitted ghost sector.
Here, the coordinates $\sigma^{(1)}(\sigma_1)$ and 
$\sigma^{(2)}(\sigma_2)$ are given by
\begin{eqnarray}
\sigma^{(1)}(\sigma_1)&=&{\alpha_1\over \alpha_1+\alpha_2}(\sigma_c+\sigma_1)
-\pi\,{\rm sgn}(\sigma_c+\sigma_1)\,,\label{eq:sigma1}\\
\sigma^{(2)}(\sigma_2)&=&{\alpha_2\over \alpha_1+\alpha_2}
\left(\sigma_c+\sigma_2
-\pi\,{\rm sgn}(\sigma_c+\sigma_2)\right),\label{eq:sigma2}
\end{eqnarray}
for $|\sigma_c+\sigma_r|<\pi,~r=1,2$,
which represent the positions of tachyon vertices
on the boundary of the joined string $3$
specified by the overlapping condition for the 3-string vertex.
Note that the phase factor $e^{i\Theta_{12}}$ appears
as a result of the $\star$ product of closed string field theory
which is computed from the last term in eq.~(4.7) in \cite{kmw1}
using (\ref{eq:formulae}) as
\begin{eqnarray}
\label{eq:Theta12}
 \Theta_{12}&=&-{1\over 2\pi}\,p_{1i}\vartheta^{ij}p_{2j}
(\sigma^{(1)}(\sigma_1)-\sigma^{(2)}(\sigma_2))
+{1\over 2}p_{1i}\vartheta^{ij}p_{2j}\,
\epsilon(\sigma^{(1)}(\sigma_1)-\sigma^{(2)}(\sigma_2))\,,
\end{eqnarray}
where 
\begin{eqnarray}
 {\vartheta}^{ij}=(2\pi\alpha')^2\left[(g-2\pi\alpha'\theta^{-1})^{-1}
{\theta}^{-1}(g+2\pi\alpha'\theta^{-1})^{-1}\right]^{ij}
\end{eqnarray}
corresponds to the noncommutativity parameter.
In the exponent, linear terms with respect to oscillators are given by
\begin{eqnarray}
\lambda_n^{(12)}&=&{\sqrt{2\alpha'}\over n}
\left(p_1\,e^{-in\sigma^{(1)}(\sigma_1)}
+p_2\,e^{-in\sigma^{(2)}(\sigma_2)}\right)(g-2\pi\alpha'\theta^{-1})^{-1}g\,,
~~\\
\tilde{\lambda}_n^{(12)}&=&{\sqrt{2\alpha'}\over n}
\left(p_1\,e^{in\sigma^{(1)}(\sigma_1)}
+p_2\,e^{in\sigma^{(2)}(\sigma_2)}\right)(g+2\pi\alpha'\theta^{-1})^{-1}g\,.
\end{eqnarray}
The factor ${\cal N}_{12}$ is evaluated as
\begin{eqnarray}
{\cal N}_{12}&=&\lim_{L\rightarrow\infty}\left[
e^{\alpha'p_1G^{-1}_Op_1\left(
\sum_{n=1}^{|\alpha_1|L}{1\over n}-
\sum_{p=1}^{|\alpha_3|L}{1\over p}\right)}
e^{\alpha'p_2G^{-1}_Op_2\left(
\sum_{n=1}^{|\alpha_2|L}{1\over n}-
\sum_{p=1}^{|\alpha_3|L}{1\over p}\right)}
\right]\nonumber\\
&=&(-\beta)^{\alpha'p_1G^{-1}_Op_1}(1+\beta)^{\alpha'p_2G^{-1}_Op_2},
~~~(\beta={\alpha_1/\alpha_3},~\alpha_3=-\alpha_1-\alpha_2),
\end{eqnarray}
where we take cutoffs for the mode number of strings
such that they are proportional to each string length parameter
$|\alpha_r|$.
This prescription was used in \cite{kmw1}v4 in order to investigate 
the on-shell condition from idempotency
and is consistent with conformal factor of the open
string tachyon vertex \cite{kmw2}.
We can also rewrite (\ref{eq:VDstar}) as
\begin{eqnarray}
&& \hat{V}_{\theta,\sigma_c}|p_1\rangle\!\rangle_{D,\alpha_1}
\star\hat{V}_{\theta,\sigma_c}|p_2\rangle\!\rangle_{D,\alpha_2}\nonumber\\
&&=~(-\beta)^{\alpha'p_1G^{-1}_Op_1}(1+\beta)^{\alpha'p_2G^{-1}_Op_2}
\,{\det}^{-{d\over 2}}(1-(\tilde{N}^{33})^2)\\
&&~~~~\times\oint {d\sigma_1\over 2\pi}
\oint {d\sigma_2\over 2\pi}\wp\,
V_{p_1}(\sigma^{(1)}(\sigma_1))
V_{p_1}(\sigma^{(2)}(\sigma_2))
|B(F=-\theta^{-1})\rangle_{\alpha_1+\alpha_2}\,,\nonumber
\end{eqnarray}
using tachyon vertex given in (\ref{eq:tach}).
This implies that  the $\star$ product of 
$\hat{V}_{\theta,\sigma_c}|p\rangle\!\rangle_{D}$
induces conventional operator product of tachyon vertices on the Neumann
boundary state.

\subsection{Seiberg-Witten limit}

Next, we proceed the third step
to take Seiberg-Witten limit \cite{SW} of 
(\ref{eq:VDstar}) in order to obtain the deformed algebra.
In the limit
$ \alpha'\sim \varepsilon^{1\over 2}\rightarrow 0\,,~
g_{ij}\sim\varepsilon \rightarrow 0$,
the $\star$ product formula (\ref{eq:VDstar}) is simplified as
\begin{eqnarray}
\label{eq:SWstar}
&& \hat{V}_{\theta,\sigma_c}|p_1\rangle\!\rangle_{D,\alpha_1}
\star\hat{V}_{\theta,\sigma_c}|p_2\rangle\!\rangle_{D,\alpha_2}
 \sim a(p_1,p_2) \hat{V}_{\theta,\sigma_c}
|p_1+p_2\rangle\!\rangle_{D,\alpha_1+\alpha_2}\,,
\\
&&a(p_1,p_2)\equiv {\det}^{-{d\over 2}}(1-(\tilde{N}^{33})^2)
\oint {d\sigma_1\over 2\pi}
\oint {d\sigma_2\over 2\pi}\,e^{i\Theta_{12}}\,,
\end{eqnarray}
where we have estimated using $\alpha^i_{-n}=\sqrt{n}\,
\tilde{e}^i_aa^{\dagger a}_n \sim
\varepsilon^{-{1\over 2}}$ and ignored linear terms in the exponent.
We can interprete that,
in this limit, the deformed Ishibashi states:
$\hat{V}_{\theta,\sigma_c}|p\rangle\!\rangle_{D}$
form a closed algebra with respect to the $\star$ product of closed 
string field theory. 
After we drop the determinant factor, the coefficient
$a(p_1,p_2)$ can be evaluated as
\begin{eqnarray}
 a(p_1,p_2)=
{\sin (-\beta p_{1i}\theta^{ij}p_{2j})\over 
-\beta p_{1i}\theta^{ij}p_{2j}}
{\sin ((1+\beta) p_{1i}\theta^{ij}p_{2j})\over 
(1+\beta) p_{1i}\theta^{ij}p_{2j}}\,,
\end{eqnarray}
where we introduce a parameter
$\beta={-\alpha_1\over \alpha_1+\alpha_2}$
($-1<\beta<0$) which comes from the assigned $\alpha$-parameters
for string fields in the $\star$ product.
The integration intervals for $\sigma_1,\sigma_2$ are taken as
$-\pi<\sigma_c+\sigma_1<\pi$, $-\pi<\sigma_c+\sigma_2<\pi$
and we have used eqs.~(\ref{eq:sigma1}) and (\ref{eq:sigma2}).
We note that the last expression does not depend 
on the cut $\sigma_c$ in the KT operator.
This independence is caused by the level matching
projections $\wp_1,\wp_2$ in the 3-string vertex.
By taking Fourier transformation, 
the induced product is represented in the coordinate space as,
\begin{eqnarray}
f_{\alpha_1}(x)\diamond_{\beta} g_{\alpha_2}(x)
&=&f_{\alpha_1}(x){\sin(-\beta\lambda)\sin((1+\beta)\lambda)\over 
(-\beta)(1+\beta)\lambda^2}g_{\alpha_2}(x)
\label{eq:closed_star}\\
&=&
f_{\alpha_1}(x)\,
\sum_{k=0}^{\infty}{(-1)^k\lambda^{2k}\over (2k+1)!}
\sum_{l=0}^k{(1+2\beta)^{2l}\over k+1}\,g_{\alpha_2}(x)\,,~~~~
\left(
\lambda={1\over 2}
{\overleftarrow{\partial}\over \partial x^i}
\theta^{ij}
{\overrightarrow{\partial}\over \partial x^j}
\right),\nonumber
\end{eqnarray}
where we have specified $\alpha_1,\alpha_2$ for coefficient functions
because the parameter $\beta$ in the above $\diamond_{\beta}$ is 
given by their ratio. In fact, for the string fields of the form
\begin{eqnarray}
\label{eq:PhiNC}
  |\hat{\Phi}_{f_{\alpha}}(\alpha)\rangle&=&\int d^dx\,f_{\alpha}(x)
\hat{V}_{\theta,\sigma_c}|\Phi_B(x,\alpha)\rangle\,,
\end{eqnarray}
where we have included the ghost and $\alpha$ sector:
$|\Phi_B(x,\alpha)\rangle=c_0^-b_0^+|B(x)\rangle\otimes
|B\rangle_{\rm ghost}\otimes|\alpha\rangle$
and $\alpha$ dependence in the coefficient function,
we can express the above $\diamond_{\beta}$ product in
terms of the $\star$ product:
\begin{eqnarray}
 &&\langle x, \alpha_1+\alpha_2|c_{-1}\tilde{c}_{-1}
|\hat{\Phi}_{f_{\alpha_1}}(\alpha_1)\rangle\star
|\hat{\Phi}_{g_{\alpha_2}}(\alpha_2)\rangle
\nonumber\\
&&=~[\mu^2\,{\det}^{-{d-2\over 2}}(1-(\tilde{N}^{33})^2)\,2\pi\delta(0)]\,
f_{\alpha_1}(x)\diamond_{\beta} g_{\alpha_2}(x)\label{eq:star12SW}
\end{eqnarray}
in the Seiberg-Witten limit.
Here, we give some comments 
on this $\diamond_{\beta}$ product (\ref{eq:closed_star}).
It is commutative in the sense:
\begin{eqnarray}
 f_{\alpha_1}(x)\diamond_{\beta} g_{\alpha_2}(x)
=g_{\alpha_2}(x)\diamond_{\beta} f_{\alpha_1}(x)
=f_{\alpha_1}(x)\diamond_{-1-\beta} g_{\alpha_2}(x)\,.
\end{eqnarray}
(Note that exchange of $\alpha_1\leftrightarrow \alpha_2$ corresponds to
 $\beta\leftrightarrow -1-\beta$.)
We can take the ``commutative'' background  limit
$\theta^{ij}\rightarrow 0$ :
\begin{eqnarray}
\lim_{\theta^{ij}\rightarrow 0}
f_{\alpha_1}(x)\diamond_{\beta} g_{\alpha_2}(x)
=f_{\alpha_1}(x)\,g_{\alpha_2}(x)
\end{eqnarray}
where the right hand side is ordinary product.
In the case that one of the string length parameter
$\alpha_1,\alpha_2$ equals to zero,
our product (\ref{eq:closed_star}) is reduced to the Strachan
product \cite{Strachan}:
\begin{eqnarray}
 &&\lim_{\alpha_1\rightarrow 0}
f_{\alpha_1}(x)\diamond_{\beta} g_{\alpha_2}(x)
=f_0(x)\diamond g_{\alpha_2}(x)\,,\nonumber\\
&&\lim_{\alpha_2\rightarrow 0}
f_{\alpha_1}(x)\diamond_{\beta} g_{\alpha_2}(x)
=f_{\alpha_1}(x)\diamond g_0(x)\,,\nonumber\\
&&{\rm where}~~~
f(x)\diamond g(x):=f(x){\sin\lambda\over \lambda}g(x)\,,~~~
\left(
\lambda={1\over 2}
{\overleftarrow{\partial}\over \partial x^i}
\theta^{ij}
{\overrightarrow{\partial}\over \partial x^j}
\right),~~~~
\label{eq:strachan}
\end{eqnarray}
which is also one of the generalized star product: $*_2$ \cite{start}.

In the literature \cite{start}, Strachan product appeared
in one-loop correction to the non-commutative Yang-Mills theory.
The appearance of the similar product here may be interpreted 
naturally.  As we have seen in section 2, taking the closed string
star product of boundary states is equivalent to the degeneration
limit of the open string one loop correction. In this interpretation,
the star product we considered can be mapped to
one-loop open string diagram with one open string external lines
attached to each of the two boundaries.
It reduces to a diagram which is similar to the one
in \cite{start} in the Seiberg-Witten limit.

It will be very interesting to obtain the explicit
form of the projector to the Strachan product,
\begin{equation}
 f\diamond f=f,
\end{equation}
since it may describe the zero-mode part of the Cardy state
that corresponds to GMS soliton. 
One important task before proceeding that direction may be,
however, to construct the argument which is valid without taking
the Seiberg-Witten limit.

We comment that the observation made here is parallel to
the situation in  {\it open} string field theory.
In the limit of a large $B$-field,
Witten's star product factorizes into that of
zero mode and nonzero modes.
The star product is then reduced to Moyal product on the zero mode sector.
The noncommutativity appears as the coefficient 
functions on the lump solution $|{\cal S}\rangle$ in the context of
vacuum string field theory \cite{openGMS}.
The correspondence is:
\begin{eqnarray}
 |{\cal S}\rangle~&\leftrightarrow &~\hat V_{\theta,\sigma_c}|B(x)\rangle
  \nonumber\\
 \mbox{Moyal product}& \leftrightarrow & \mbox{Strachan product}
  \nonumber\\
\mbox{Open string field theory}&\leftrightarrow &
\mbox{Closed string field theory}\,.\nonumber
\end{eqnarray}
This may be a natural extension
of open vs. closed ``VSFT'' correspondence,
as we suggested in \cite{kmw1, kmw2}, 
for a constant $B$-field background in the transverse directions.

\section{Conclusion and Discussion}

A main observation in this article is that the nonlinear relation
(\ref{e_idempotency}) is satisfied by arbitrary
consistent boundary states in the sense of Cardy
for any conformal invariant background.
The origin of such a simple relation is the factorization
property of the boundary conformal field theory.
Since this should be true for any background as an axiom,
our nonlinear equation should be true for any consistent
closed string field theory.
In fact, we have checked this relation
for torus and $Z_2$ orbifold 
by direct calculation in terms of 
explicit oscillator formulation of the HIKKO
closed string field theory.

Although the relation (\ref{e_idempotency}) looks exactly like
a VSFT equation, it is not a consequence of a particular proposal
of the closed string field theory.  Usually it is believed that
such an equation for the vacuum theory can be obtained from
the re-expansion around  the tachyon vacuum
of some consistent string field theory. However, our equation
is not, at least at present, obtained in that way.
It is rather a direct consequence of an axiom of 
the boundary conformal field theory.  
It is very interesting that a universal nonlinear equation
can be obtained in this way.
In a sense, it is more like loop equation.

A weak point of our equation may be that it contains the regularization
parameter $K$ explicitly and divergent while it is milder
for the HIKKO type vertex than Zwiebach's one.
This can be overcome by the generalization
to superstring field theory.
The factorization property of two holes attached
to a BPS D-brane is regular since the open string channel 
does not contain tachyon.  In this sense, it will be possible to
write down a regular nonlinear equation which characterize the
BPS D-branes.
A complication arises when we consider a product of non-BPS D-brane
or different type of BPS D-branes.  
In such a situation,
there appears the open string tachyon and their star product will be
divergent.  This will be very different from the bosonic case
where eq.~(\ref{e_idempotency}) is universally true for any D-brane.
We will come back to this question in our future study.

\section*{Acknowledgements}

We are obliged to
H.~Isono and E.~Watanabe for helpful discussions.
I.~K. would like to thank H.~Kunitomo
for valuable comments.
I.~K. would also like to thank
the Yukawa Institute for Theoretical Physics at Kyoto
University. Discussions during the YITP workshop YITP-W-04-03 on
``Quantum Field Theory 2004'' were useful to complete this work.
Y.~M would like to thank G.~Semenoff for the invitation
to a workshop ``String Field Theory Camp'' at BIRS where very useful
discussions with the participants were possible.  
He would like to thank, especially, S.~Minwalla and W.~Taylor
for their comments and interests.
I.~K. is supported in part by JSPS Research Fellowships
for Young Scientists. Y.~M. is supported in part by Grant-in-Aid (\#
16540232) from the Ministry of Education, Science, Sports and Culture of
Japan.

\appendix

\section{Star product on $Z_2$ orbifold
\label{sec:starproduct}}

In this section, we briefly review the star product on $T^D/{\bf Z}_2$
orbifold \cite{Itoh_Kunitomo} and fix our convention which is mainly
based on \cite{KZ}.
By restricting to the untwisted sector and removing ${\bf Z}_2$ projection,
we obtain the star product on a torus $T^D$.

We define the $\star$ product for the string fields 
$|A\rangle,|B\rangle$ by:
\begin{eqnarray}
&&|A\rangle \star |B\rangle\equiv |A\star B\rangle_3
\equiv~{}_1\langle A|~ {}_2\langle B|V(1,2,3)\rangle\,,\\
&&~~{\rm where}~~~
{}_2\langle \Phi|\equiv \langle R(1,2)|\Phi\rangle_1\,,
\end{eqnarray}
which gives cubic interaction term in an action of closed string
filed theory.
In order to define the above concretely,
 we should specify the reflector $\langle
R(1,2)|$ and the 3-string vertex $|V(1,2,3)\rangle$
in $T^D/{\bf Z}_2$ sector. 
We expand the coordinates $X^i(\sigma)$ and their canonical conjugate
momentum $P_i(\sigma)$ to express them in terms of oscillators as follows.
In the untwisted sector, $X^i(\sigma+2\pi)\equiv X^i(\sigma)~({\rm
mod}~2\pi\sqrt{\alpha'})$:
\begin{eqnarray}
 \label{eq:mode_st}
 X^i(\sigma)&=&\sqrt{\alpha'}[x^i+w^i\sigma]
+i\sqrt{\alpha'\over 2}\sum_{n\ne 0,\,n\in {\bf Z}}{1\over n}
\left[\alpha_n^ie^{in\sigma}
+\tilde{\alpha}_n^ie^{-in\sigma}\right],~~~~\\
P_i(\sigma)&=&{1\over 2\pi\sqrt{\alpha'}}\left[
p_i+{1\over\sqrt{2}}\sum_{n\ne 0,\,n\in {\bf Z}}\left(E^T_{ij}\alpha_n^j
e^{in\sigma}+E_{ij}\tilde{\alpha}_n^je^{-in\sigma}\right)
\right],
\end{eqnarray}
where $E_{ij}=G_{ij}+2\pi\alpha' B_{ij},E^T_{ij}=G_{ij}-2\pi\alpha'
B_{ij}$. 
The commutation relations are given by
$
[x^i,p_j]=i\delta^i_j,[\alpha^i_n,\alpha^j_m]=nG^{ij}\delta_{n+m,0},
[\tilde{\alpha}^i_n,\tilde{\alpha}^j_m]=nG^{ij}\delta_{n+m,0}.
$
In our compactification, we should identify as $x^i\equiv x^i+2\pi$
and then the zero mode momentum $p_i$ takes integer eigenvalue.
In the twisted sector, $X^i(\sigma+2\pi)\equiv -X^i(\sigma)~({\rm
mod}~2\pi\sqrt{\alpha'})$:
\begin{eqnarray}
 \label{eq:t_mode_st}
 X^i(\sigma)&=&\sqrt{\alpha'}\,x^i
+i\sqrt{\alpha'\over 2}\sum_{r\in {\bf Z}+{1\over 2}}{1\over r}
\left[\alpha_r^ie^{ir\sigma}
+\tilde{\alpha}_r^i e^{-ir\sigma}\right],\\
P_i(\sigma)&=&{1\over 2\pi\sqrt{2\alpha'}}
\sum_{r\in {\bf Z}+{1\over 2}}\left(E^T_{ij}\alpha_r^j
e^{ir\sigma}+E_{ij}\tilde{\alpha}_r^je^{-ir\sigma}\right).
\end{eqnarray}
The commutation relations of nonzero modes are given by
$[\alpha^i_r,\alpha^j_s]=rG^{ij}\delta_{r+s,0}$, 
$[\tilde{\alpha}^i_r,\tilde{\alpha}^j_s]=rG^{ij}\delta_{r+s,0}$
and the zero mode $x^i$ takes eigenvalue corresponding to fixed points
of ${\bf Z}_2$ action: $x^i=\pi (n^f)^i$ where $(n^f)^i=0$ or $1$.

\paragraph{Reflector}
We use reflector to obtain a bra $\langle \Phi|$ from a ket $|\Phi\rangle$.
There are two types of reflector according to the twisted/untwisted sector.
For the untwisted sector,\footnote{
We often denote 
$\wp_1\cdots \wp_N$ as
$\wp_{1\cdots N}$ where $\wp_r$ is a projector which
imposes the level matching condition
$L_0^{(r)}-\tilde{L}_0^{(r)}=0$ on each string field.
}
\begin{eqnarray}
\label{eq:Ruu}
 \langle R_u(1,2)|&=&
\sum_{p_r,w_r}\delta^D_{p_1+p_2,0}\delta^D_{w_1+w_2,0}
\langle p_1,w_1|\langle p_2,w_2|\,
e^{E_u(1,2)}e^{-i\pi p_1w_1}\wp_{12},~~~~\\
E_u(1,2)&=&-\sum_{n\ge 1}  {(-1)^n\over n}
G_{ij}\left(\alpha_n^{(1)i} \alpha_n^{(2)j}+
\tilde{\alpha}_n^{(1)i} \tilde{\alpha}_n^{(2)j} \right)\,,~
\end{eqnarray}
where the the prefactor $e^{-i\pi p_1w_1}$ comes from the connection
condition $X^{(1)}(\sigma)-X^{(2)}(\pi-\sigma)=0$
 without projector $\wp_{12}$ \cite{KZ}\footnote{
This factor should be removed if we remove $(-1)^n$ in $E_u(1,2)$
and this implies a different connection condition 
$X^{(1)}(\sigma)-X^{(2)}(-\sigma)=0$ without $\wp_{12}$.
By multiplying $\wp_{12}$, these two conventions become equivalent
for the reflector $ \langle R_u(1,2)|$.
} and the oscillator vacuum with zero mode eigen value $(p_i,w^i)$:
 $\langle p,w|$ is normalized as
$\langle p,w|p',w'\rangle=\delta^D_{p,p'}\delta^D_{w,w'}$.
For the twisted sector, the reflector is given by
\begin{eqnarray}
\label{eq:Rtt}
 \langle R_t(1,2)|&=&
\sum_{n_1^f,\,n_2^f}\delta^D_{n_1^f,\,n_2^f}
\,\langle n_1^f|\langle n_2^f|\,e^{-\sum_{r\ge {1\over 2}}{1\over r}
G_{ij}\left(\alpha_r^{(1)i} \alpha_r^{(2)j}+
\tilde{\alpha}_r^{(1)i} \tilde{\alpha}_r^{(2)j} \right)
}\wp_{12},~~~
\end{eqnarray}
which represents $X^{(1)}(\sigma)-X^{(2)}(-\sigma)=0$ without
$\wp_{12}$ and we take the normalization of the oscillator vacuum 
for the fixed point $\pi n^f$ as $\langle
n^f|n^{f\prime}\rangle=\delta^D_{n^f,n^{f\prime}}$.

\paragraph{3-string vertex}
We have two types of 3-string interaction:
(uuu) all strings are in the untwisted sector; (utt)
one is in the untwisted sector and the other two are in the 
twisted sector.
Correspondingly, there are two types of 3-string vertex.
They are constructed by a connection condition based on 
HIKKO type interaction, i.e., joining/splitting of closed strings at one
interaction point.
(Odd number of twisted sectors such as (ttt), (uut) are not contained
in 3-string interaction terms to be consistent with ${\bf Z}_2$ action.)

For (uuu)-type 3-string vertex, by assigning $\alpha_r$ for each string,
 we have
\begin{eqnarray}
\label{eq:Vuuu}
 |V(1_u,2_u,3_u)\rangle
&=&\wp_{123}{\cal P}_{u1}^{Z_2}{\cal P}_{u2}^{Z_2}{\cal P}_{u3}^{Z_2}
\sum_{p_r,w_r}\delta_{p_1+p_2+p_3,0}\delta_{w_1+w_2+w_3,0}
\nonumber\\
&&\times\, e^{-i\pi(p_3w_2-p_1w_1)}e^{E_u(1,2,3)}
|p_1,w_1\rangle|p_2,w_2\rangle|p_3,w_3\rangle,~~~~~
\end{eqnarray}
where the exponent is given by
\begin{eqnarray}
&&E_u(1,2,3)
={1\over 2}\sum_{r,s=1}^3\sum_{n,m\ge 0}
\bar{N}^{rs}_{nm}G_{ij}\left(\alpha_{-n}^{i(r)}
\alpha_{-m}^{j(s)}+\tilde{\alpha}_{-n}^{i(r)}
\tilde{\alpha}_{-m}^{j(s)}\right)\,.
\end{eqnarray}
Here $\bar{N}^{rs}_{nm}$ is the same as the Neumann
coefficient on ${\bf R}^d$ (we also use the notation:
$\tilde{N}^{rs}_{nm}:=\sqrt{nm}\bar{N}^{rs}_{nm}\,~(n,m>0)$) \cite{HIKKO2}
and we define zero modes as:
$\alpha_0^i=G^{ij}(p_j-E_{jk}w^k)/\sqrt{2}$,
$\tilde{\alpha}_0^i=G^{ij}(p_j+E_{jk}^Tw^k)/\sqrt{2}$.
The prefactor ${\cal P}_{u}^{Z_2}$ is ${\bf Z}_2$-projection for the
untwisted 
sector and is given by ${\cal P}_u^{Z_2}={1\over 2}(1+RO_u)$ with
$R|p,w\rangle=|-p,-w\rangle\,,~O_u\alpha_n^iO^{-1}_u=-\alpha_n^i\,,
~O_u\tilde{\alpha}_n^iO^{-1}_u=-\tilde{\alpha}_n^i$.
The phase factor $e^{-i\pi(p_3w_2-p_1w_1)}$ is necessary to satisfy
Jacobi identity \cite{HIKKO_torus, Maeno_Takano}.
The above vertex $|V(1_u,2_u,3_u)\rangle$
 is also obtained by multiplying ${\bf Z}_2$-projection 
${\cal P}_{u1}^{Z_2}{\cal P}_{u2}^{Z_2}{\cal P}_{u3}^{Z_2}$ 
to the 3-string vertex on the torus $T^D$ \cite{HIKKO_torus}.

For (utt)-type 3-string vertex, by assigning $\alpha_r$ for each string,
 we have 
\begin{eqnarray}
\label{eq:Vutt}
 |V(1_u,2_t,3_t)\rangle&=&
e^{{D\over 8}\tau_0\left(\alpha_2^{-1}+\alpha_3^{-1}\right)}\wp_{123}
{\cal P}_{u1}^{Z_2}{\cal P}_{t2}^{Z_2}{\cal P}_{t3}^{Z_2}\\
&&\times\,
\sum_{p_1,w_1}\sum_{n_2^f,\,n_3^f}
\gamma({\bf p}_1;n_2^f,n_3^f)
e^{E_t(1_t,2_u,3_u)}|p_1,w_1\rangle|n_2^f\rangle|n_3^f\rangle\,,\nonumber\\
E_t(1_t,2_u,3_u)&=&{1\over 2}\sum_{r,s=1}^3\sum_{n_r,m_s\ge 0}
T^{rs}_{n_r m_s}G_{ij}\left(\alpha_{-n_r}^{i(r)}
\alpha_{-m_s}^{j(s)}+\tilde{\alpha}_{-n_r}^{i(r)}
\tilde{\alpha}_{-m_s}^{j(s)}\right),~~~~~~
\end{eqnarray}
where Neumann coefficients $T^{rs}_{n_rm_s}$ are given explicitly 
in Appendix \ref{sec:TNeumann} and
\begin{eqnarray}
\gamma({\bf p}_1;n_2^f,n_3^f) 
&=&(-1)^{p_1n_3^f}\,\sum_{m^i\in {\bf Z}}\delta^D_{n_3^f-n_2^f+w_1+2m,0}
\end{eqnarray}
is the cocycle factor \cite{Itoh_Kunitomo, IKKS} and 
${\cal P}_{t}^{Z_2}={1\over 2}(1+O_t),$ which is given by
$O_t\alpha_r^i O^{-1}_t=-\alpha_r^i,
O_t\tilde{\alpha}_r^i O^{-1}_t=-\tilde{\alpha}_r^i$,
is  the ${\bf Z}_2$-projection.
The extra factor $e^{{D\over
8}\tau_0\left(\alpha_2^{-1}+\alpha_3^{-1}\right)}$,
($\tau_0=\sum_{r=1}^3\alpha_r\log|\alpha_r|$,
$\alpha_1+\alpha_2+\alpha_3=0$),
can be identified with the conformal factor of twist fields
in CFT language.

Note that the complete 3-string vertex is given by 
including ghost, matter ${\bf R}^d$ and $\alpha$ sector in the above
expression (\ref{eq:Vuuu}) or (\ref{eq:Vutt}).

\section{Neumann coefficients for the 
twisted sector on ${\bf Z}_2$ orbifold
\label{sec:TNeumann}}

The Neumann coefficients $T^{rs}_{n_rm_s}$ in (\ref{eq:Vutt})
are given by
$T^{11}_{00}=-2\log 2+{\tau_0\over \alpha_1}$\, 
and integration form in \cite{Itoh_Kunitomo}.
We can demonstrate that there is a relation:
\begin{equation}
\label{eq:Trs}
   T^{rs}_{n_r m_s}={\alpha_1n_rm_s\over \alpha_rm_s+\alpha_sn_r}
\,T^{r1}_{n_r0}\,T^{s1}_{m_s0}\,,~~~(n_r,m_r>0)\,,
\end{equation}
and $T^{r1}_{n_r0}$ are explicitly obtained:
\begin{eqnarray}
T^{11}_{n0}&=&{e^{n{\tau_0\over \alpha_1}}\over n}
{\Gamma\left({1\over 2}-{\alpha_2\over \alpha_1}n\right)\over 
n!\,\Gamma\left({1\over 2}+{\alpha_3\over\alpha_1}n\right)}
\,,~~~n=1,2,\cdots,\\
T^{21}_{r0}&=&{e^{r{\tau_0\over \alpha_2}}\over r}
{(-1)^{r+{1\over 2}}\,\Gamma\left(-{\alpha_1\over \alpha_2}r\right)
\over \left(r-{1\over 2}\right)!\,
\Gamma\left({1\over 2}+{\alpha_3\over \alpha_2}r\right)}\,,~~~
r={1\over 2},{3\over 2},\cdots,\\
T^{31}_{r0}&=&{e^{r{\tau_0\over \alpha_3}}\over r}
{(-1)^{r+{1\over 2}}\,\Gamma\left(-{\alpha_1\over \alpha_3}r\right)
\over \left(r-{1\over 2}\right)!\,
\Gamma\left({1\over 2}+{\alpha_2\over \alpha_3}r\right)}\,,~~~
r={1\over 2},{3\over 2},\cdots.
\end{eqnarray}
Note that only string 1 is in the untwisted sector which includes zero
mode $(p,w)$
in the (utt) type 3-string vertex (\ref{eq:Vutt}). However, this
structure of the Neumann coefficients $T^{rs}_{n_rm_s}$ is similar to 
that of $\bar{N}^{rs}_{nm}$ \cite{Mandel64}
 in the untwisted 3-string vertex
(\ref{eq:Vuuu}) in which all 3 strings have zero mode $(p,w)$.

Using continuity of Neumann function $T(\rho,\tilde{\rho})$ which is
given in \cite{Itoh_Kunitomo}
with the  method in Appendix B in \cite{Kawano_Takahashi},
namely, from the identity $\sum_{t=1}^3\int_{-\pi}^{\pi}d\sigma_t^{\prime}
T(\rho_r,\rho_t^{\prime})
{\partial\over \partial\xi_t^{\prime}}T(\rho_t^{\prime},\rho_s^{\prime\prime})
=0$ (where ${\rm Re}\,\rho'_t=\tau_0$), we have obtained the relations:
\begin{eqnarray}
&& \sum_{t=1}^3\sum_{l_t> 0}T^{rt}_{n_r l_t}\, l_t\, T^{ts}_{l_t m_s}
=\delta_{r,s}\delta_{n_r,m_s}{1\over n_r}\,,\nonumber\\
&& \sum_{t=1}^3\sum_{l_t> 0}T^{1t}_{0 l_t}\, l_t\, T^{ts}_{l_t m_s}
=-T^{1s}_{0 m_s}\,,~~~~~~~~~
 \sum_{t=1}^3\sum_{l_t> 0}T^{1t}_{0 l_t}\, l_t\, T^{t1}_{l_t 0}
=-2T^{11}_{00}\,,~~~
\label{eq:Yoneya}
\end{eqnarray}
which correspond to Yoneya formulae for the untwisted sector
\cite{Yoneya}.
These are essential to simplify some expressions in terms of 
Neumann coefficients which appear in computation of the $\star$ product.

Furthermore, in the case of $\alpha_1>0,\alpha_2<0,\alpha_3<0$, we
can derive following formulae using the method in \cite{GS}:
\begin{eqnarray}
 \tilde{T}_{n_rn_s}^{rs}&:=&\sqrt{n_rm_s}\,T_{n_rn_s}^{rs}=
(\delta_{r,s}1-2A^{\prime(r)T}\Gamma^{\prime-1}A^{\prime(s)})_{n_rn_s}
~~~(n_r,n_s>0)\nonumber\\
&=&{\alpha_1n_rn_s\over n_s\alpha_r+n_r\alpha_s}
(A^{\prime(r)T}\Gamma^{\prime-1}B^{\prime})_{n_r}
(A^{\prime(s)T}\Gamma^{\prime-1}B^{\prime})_{n_s}\,,\\
\tilde{T}^{r1}_{n_r0}&:=&\sqrt{n_r}\,T^{r1}_{n_r0}=
(A^{\prime(r)T}\Gamma^{\prime-1}B^{\prime})_{n_r}\,,
~~~~~~~~~~~~~~~~~~(n_r>0)\\
T^{11}_{00}&=&-{1\over 2}B^{\prime T}\Gamma^{\prime-1}B^{\prime}\,,
\end{eqnarray}
where the infinite
 matrices $A^{\prime(r)}_{nm_r},\Gamma^{\prime}_{nm}$ and the 
infinite vector $B^{\prime}_n$ are given by
\begin{eqnarray}
&&A^{\prime(r)}_{nm_r}=(-1)^{n+m_r-{1\over 2}}
{2n^{3\over 2}
\left(\alpha_r\over \alpha_1\right)^2
\cos \left({\alpha_r\over \alpha_1}n\pi\right)
\over \pi m_r^{1\over 2}
\left[m_r^2-n^2\left(\alpha_r\over \alpha_1\right)^2\right]}
\,,~~~r=2,3;~~m_r\ge {1\over 2}\,,~~~~~~\\
&&A^{\prime(1)}_{nm}=\delta_{n,m}\,,~~~~
B^{\prime}_n={2(-1)^n\cos n\pi\beta\over \sqrt{n}}\,,~~~
~~~n,m\ge 1\,,\\
&&\Gamma^{\prime}_{nm}=\sum_{r=1}^3\sum_{l_r>0}
A^{\prime(r)}_{nl_r}A^{\prime(r)}_{ml_r}\,,~~~~~
n,m\ge 1\,.
\end{eqnarray}
Using these formulae,
we can prove various identities, including (\ref{eq:Yoneya}),
which correspond to those in \cite{GS} such as
\begin{eqnarray}
&&\sum_{l\ge 1}A^{\prime(r)}_{ln_r}{1\over l}A^{\prime(s)}_{ln_s}
=-{\alpha_r\over n_r\alpha_1}\delta_{r,s}\,,~~~~~r,s=2,3\,,\\
&&\sum_{r=2,3}\sum_{l_r\ge{1\over 2}}A^{\prime(r)}_{ml_r}
{l_r\over \alpha_r}
A^{\prime(r)}_{nl_r}=-{m\over \alpha_1}\delta_{m,n}\,,~~~~~~
m,n\ge 1\,,\\
&&\sum_{r=1}^3\sum_{l_r>0}{\alpha_r\over \alpha_1}A^{\prime(r)}_{nl_r}
{1\over l_r}A^{\prime(r)}_{ml_r}
=-{1\over 2}B^{\prime}_nB^{\prime}_m\,,~~~~~~
m,n\ge 1\,.
\end{eqnarray}

\section{Cremmer-Gervais identity for $T^{rs}_{n_rm_s}$
\label{sec:CG}}

\begin{figure}[htbp]
	\begin{center}
	\scalebox{0.6}[0.6]{\includegraphics{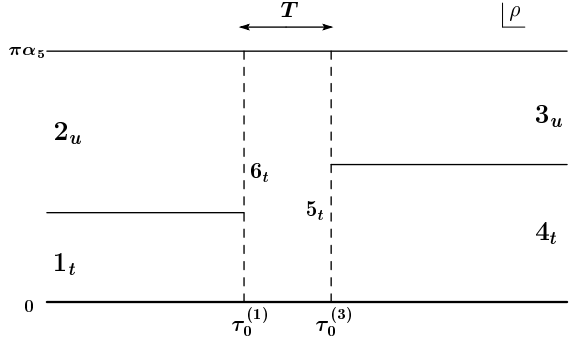}}
	\end{center}
	\caption{
4-string configuration in $\rho$-plane.
We have drawn ${\rm Im}\,\rho\ge 0$ only.
We take  strings 1,4 (2,3) 
in the twisted (untwisted) sector.
The intermediate strings 6,5 are in the twisted sector.
(There is a ${\bf Z}_2$ cut at ${\rm Im}\,\rho=0$.)}
	\label{fig:CG(P)}
\end{figure}

We demonstrate the relation (\ref{eq:CG_T}) by using 
an analogue of Cremmer-Gervais identity \cite{CG}.
Let us consider  matrices such as
\begin{eqnarray}
\label{eq:AB}
&& {\tilde{{\cal N}}^{66}}_{nm}={nm\over n+m}A_{n}A_{m}\,,~~~~
{\tilde{{\cal N}}^{55}}_{t\,nm}={nm\over n+m}B_{n}B_{m}
e^{-(n+m)t}\,,~~~~~~
\end{eqnarray}
which are the same form as the Neumann matrix $\tilde{N}^{rr}$
for 3-string vertex in the untwisted sector
and its $T=|\alpha_5| t$ evolved one.
We can derive a differential equation:
\begin{eqnarray}
\label{eq:CG_diffeq}
&&{\partial^2\over \partial t^2}
\log  \det(1-{\tilde{\cal{N}}^{66}}{\tilde{\cal N}_t}^{55})=
-{1\over 4}\left(\partial_t^2a_{00}\over \partial_t b_{00}\right)^2\,,
\end{eqnarray}
by direct computation, where
\begin{eqnarray}
 a_{00}&=&\sum_{n,m}A_n\left(\tilde{{\cal N}}^{55}_t
(1-\tilde{{\cal N}}^{66}\tilde{\cal N}^{55}_t)^{-1}\right)_{nm}A_m\,,\\
b_{00}&=&\sum_{n,m}B_ne^{-nt}\left(
(1-\tilde{{\cal N}}^{66}\tilde{\cal N}^{55}_t)^{-1}\right)_{nm}A_m\,.
\end{eqnarray}
The counterpart of (\ref{eq:CG_diffeq}) was integrated by
identifying $a_{00},b_{00}$ with Neumann coefficients for 4-string
vertex \cite{CG}.
As we have noted in Appendix \ref{sec:TNeumann}, the Neumann matrices 
for the twisted sector $\tilde{T}^{rs}$ also has  the same structure.
Therefore, we consider the replacement in (\ref{eq:AB}):
\begin{eqnarray}
\label{eq:repl}
 A_n,~~B_n,~~t~~~\longrightarrow~~~
\left(\alpha_2/\alpha_6\right)^{1\over 2}\tilde{T}^{62}_{n0},~~
\left(\alpha_3/\alpha_5\right)^{1\over 2}\tilde{T}^{53}_{n0},~~
T/\alpha_5\,,
\end{eqnarray}
respectively, to evaluate the determinant
$\det(1-\tilde{T}^{6_t6_t}\tilde{T}^{5_t5_t}_t)$.
We have depicted this situation in Fig.~\ref{fig:CG(P)}.
In particular strings 2 and 3 are in the untwisted sector.
The Neumann coefficients for this 4-string amplitude 
can be obtained by expanding the Neumann function 
\begin{eqnarray}
&&{\cal T}(\rho,\tilde{\rho})=\log\!\left[\!\sqrt{z-Z_4\over z-Z_1}
-\sqrt{\tilde{z}-Z_4\over \tilde{z}-Z_1}\right]\!
-
\log\!\left[\!\sqrt{z-Z_4\over z-Z_1}\!
+\!\sqrt{\tilde{z}-Z_4\over \tilde{z}-Z_1}\right],~~~~~~~
\end{eqnarray}
with the Mandelstam mapping:
$ \rho(z)=\sum_{r=1}^4\alpha_r\log(z-Z_r)
=\alpha_r\zeta_r+\tau^{(r)}_0+i\beta_r$,
where $\tau_0^{(1)}=\tau_0^{(2)}={\rm Re}\,\rho(z_-),\,
\tau_0^{(3)}=\tau_0^{(4)}={\rm Re}\,\rho(z_+)$ are interaction
time: $\left.{d\rho\over dz}\right|_{z_{\pm}}=0$ (Fig.~\ref{fig:CG(P)}).
This procedure is parallel to that for constructing
3-string vertex (\ref{eq:Vutt}) in \cite{Itoh_Kunitomo}.
In particular, the coefficient for zero modes are obtained as
\begin{eqnarray}
 T^{(4)rs}_{~~00}&=&\log\!\left[\!\sqrt{Z_r-Z_4\over Z_r-Z_1}
\!-\sqrt{Z_s-Z_4\over Z_s-Z_1}\right]\!
-
\log\!\left[\!\sqrt{Z_r-Z_4\over Z_r-Z_1}\!
+\!\sqrt{Z_s-Z_4\over Z_s-Z_1}\right],~~\nonumber\\
&&~~~~~~~~(r,s=2,3,~r\ne s),~~\\
T^{(4)rr}_{~~00}&=&-2\log 2+{\tau_0^{(r)}+i\beta_r\over \alpha_r}
-\log(Z_r-Z_1)-\log(Z_r-Z_4)\nonumber\\
&&+\log(Z_4-Z_1)
-\sum_{l\ne r,l=1,\cdots,4}{\alpha_l\over \alpha_r}\log(Z_r-Z_l),
~~(r=2,3)\,.
\end{eqnarray}
By comparing them with zero mode dependence in the exponent of
\begin{eqnarray}
 \langle R(5_t,6_t)|e^{-{T\over \alpha_5}(L_0^{(5)}+{\tilde{L}_0^{(5)}})}
|V_0(1_t,2_u,6_t)\rangle|V_{0}(5_t,3_u,4_t)\rangle\,,
\end{eqnarray}
(where $\langle R(5_t,6_t)|$ and 
$|V_0(r,s,t)\rangle$ are reflector (\ref{eq:Rtt}) and 
3-string vertex (\ref{eq:Vutt}) respectively,
with appropriate replacement and without projections)
which represents Fig.~\ref{fig:CG(P)}, we can 
make an identification:
\begin{eqnarray}
\label{eq:T4ab}
T^{(4)22}_{~~00}
=-{\alpha_5\over \alpha_2}a_{00}+T^{22}_{00}\,,~~~~
T^{(4)23}_{~~00}
=-{\alpha_5\over \sqrt{-\alpha_2\alpha_3}}\,b_{00}\,,
\end{eqnarray}
up to pure imaginary constant
where $
 \alpha_1+\alpha_2+\alpha_3+\alpha_4=0\,,~\alpha_3+\alpha_4+\alpha_5=0\,
$.
By fixing as $Z_1=\infty,Z_2=1,Z_4=0$, we have some relations:
\begin{eqnarray}
 z_{\pm}&=&-(2\alpha_1)^{-1}(\alpha_{34}+\alpha_{24}Z_3\pm
\Delta^{1\over 2})\,,~~~~(\alpha_{ij}:=\alpha_i+\alpha_j)\,,\\
\Delta&=&\alpha_{24}^2Z_3^2+2(\alpha_2\alpha_3+\alpha_4\alpha_1)Z_3
+\alpha_{34}^2\,,\\
{\partial T\over \partial Z_3}&=&-{\Delta^{1\over 2}\over Z_3(1-Z_3)}
\,,~~~
T:=\tau_0^{(3)}-\tau_0^{(1)}\,,
\end{eqnarray}
which are the same convention as in \cite{HIKKO2} Appendix C, and
then we obtain a differential equation for determinant of Neumann
coefficients with regularization parameter $T$:
\begin{eqnarray}
\label{eq:det(P)}
&& {\partial^2\over \partial T^2}\log
\det(1-\tilde{T}^{66}\tilde{T}^{55}_t)=
{\alpha_2\alpha_3(\alpha_5-(\alpha_2+\alpha_4)Z_3)^2Z_3(1-Z_3)^2\over 
4\Delta^3},~~~~~~~~
\end{eqnarray}
using (\ref{eq:CG_diffeq}),(\ref{eq:repl}) and (\ref{eq:T4ab}).
This can be rewritten by subtracting the counterpart in the untwisted sector
as:
\begin{eqnarray}
 && {\partial^2\over \partial T^2}\log\left[
\det(1-\tilde{T}^{66}\tilde{T}^{55}_t)\over 
\det(1-\tilde{N}^{66}\tilde{N}^{55}_t)
\right]
={\partial^2\over \partial T^2}\left[
-{1\over 4}\,a_{00}+{\alpha_2\alpha_5(\alpha_1-\alpha_4)\over 8\alpha_4}\,a
\right]\!,~~~~~~~~
\end{eqnarray}
where $a$ is given in (C.18) \cite{HIKKO2}.
Around $Z_3\sim 0$, we can estimate these determinants:
$
\log\det(1-\tilde{T}^{66}\tilde{T}^{55}_t)={\cal O}(Z_3),
\log\det(1-\tilde{N}^{66}\tilde{N}^{55}_t)={\cal O}(Z_3^2)
$
by definition. Therefore, we have obtained:
\begin{eqnarray}
&&\log\left[
\det(1-\tilde{T}^{66}\tilde{T}^{55}_t)\over 
\det(1-\tilde{N}^{66}\tilde{N}^{55}_t)
\right]\\
&&={\alpha_4\alpha_{34}-\alpha_1(\alpha_3-\alpha_4)
\over 16\alpha_1\alpha_4}\!\left[
{\tau^{534}_0-\tau^{126}_0-T\over \alpha_5}-\log Z_3\right]\!
+\!{\alpha_{14}^2\over 16\alpha_1\alpha_4}\log(1-Z_3),\nonumber
\end{eqnarray}
up to pure imaginary constant, where 
$\tau_0^{ijk}:=\sum_{r=i,j,k}\alpha_r\log|\alpha_r|$.
In order to evaluate the ratio of left and right hand side in
(\ref{eq:CG_T})  by regularizing the Neumann matrices 
with $T$ such as Appendix B in \cite{kmw2},
we take  $\alpha_3=-\alpha_2,\alpha_4=-\alpha_1$ in particular, and 
we get
\begin{eqnarray}
&&\log\left|
{e^{{1\over 8}\tau_0(\alpha_2^{-1}-(\alpha_1+\alpha_2)^{-1})}
(\det(1-\tilde{T}^{66}\tilde{T}^{55}_t))^{-{1\over 2}}\over
(\det(1-\tilde{N}^{66}\tilde{N}^{55}_t))^{-{1\over 2}}}
\right|
=
-{\alpha_2\over 16\alpha_1}\left({T\over \alpha_1+\alpha_2}+
\log Z_3\right)
={\cal O}(T)\,.~~~~~~~~
\end{eqnarray}
(Note that $T\sim 0$ corresponds to $Z_3\sim 1$.)
This implies the relation (\ref{eq:CG_T}) for $T\rightarrow +0$.

\section{Evaluation of the coefficient $c_t$
\label{sec:1-loop}}

Using the similar method in Appendix \ref{sec:CG},
we cannot evaluate $c_t$ (\ref{eq:ct}) because 
the counterpart in Fig.~\ref{fig:CG(P)} is 4-twisted string
and we cannot refer to $T^{(4)rs}_{~~00}$ in order to solve a differential
equation such as (\ref{eq:CG_diffeq}).
Therefore, we consider a different regularization such as \S \ref{sec:compC}.
Using (\ref{eq:tt_D}),(\ref{eq:Bnf}) and 
(\ref{eq:Vutt}),
the determinant of Neumann coefficients is
represented as:
\begin{eqnarray}
\label{eq:CDp}
{\cal C}'_D&:=&e^{{D\over 8}\tau_0(\alpha_1^{-1}+\alpha_2^{-1})}
{\det}^{-{D\over 2}}(1-(\tilde{T}^{3_u3_u})^2)\nonumber\\
&=&
{}_{\alpha_1+\alpha_2}\langle p=0,w=0 |\,(\,|B_{n^f}
\rangle_{t,\alpha_1}\star|B_{n^f}\rangle_{t,\alpha_2})\,.
\end{eqnarray}
We regularize ${\cal C}'_D$ by inserting $e^{-{\tau_1\over
2\alpha_r}(L_0+\tilde{L}_0-2a_t)}$ in front of
$|B_{n^f}\rangle_{t,\alpha_r}$ ($r=1,2$) 
where $L_0-a_t=\sum_{r\ge 1/2}\alpha_{-r}^iG_{ij}\alpha_r^j+{D\over 48}$
and $\tilde{L}_0-a_t=\sum_{r\ge 1/2}\tilde{\alpha}_{-r}^iG_{ij}
\tilde{\alpha}_r^j+{D\over 48}$.
In order to evaluate ${\cal C}^{\prime}_D$ using the method 
in \S \ref{sec:compC}, we should take degenerate limit of 
\begin{eqnarray}
\label{eq:BBt}
{}_t\langle B_{n^f}|
\tilde{q}^{{1\over 2}(L_0+\tilde{L}_0-2a_t)}
|B_{n^f}\rangle_t
=\left(\eta(\tilde{\tau})\over \vartheta_0(0|\tilde{\tau})\right)^{D\over 2}
=\left(\eta(-1/\tilde{\tau})\over 
\vartheta_2(0|-1/\tilde{\tau})\right)^{D\over 2}\,,
\end{eqnarray}
which comes from evaluation of the amplitude in Fig.~\ref{fig:mapping}-b
with ${\bf Z}_2$ cut along ${\rm Re}\,u=-1/2$.
Similarly, we regularize
\begin{eqnarray}
\label{eq:CDdef}
{\cal C}_D&:=&
{\det}^{-{D\over 2}}(1-(\tilde{N}^{33})^2)\\
&=&
{}_{\alpha_1+\alpha_2}\langle p=0,w=0 |(|B_{n^f}
\rangle_{u,\alpha_1}\star|B_{n^f}\rangle_{u,\alpha_2})(2\pi\delta(0))^{-D}
{\det}^{1\over 2}(2G)\,,\nonumber
\end{eqnarray}
(which follows from (\ref{eq:uu_D}))
and evaluate it by taking degenerate limit of
\begin{eqnarray}
\label{eq:BBu}
&& (2\pi\delta(0))^{-D}{\det}^{1\over 2}(2G)\,{}_u\langle B_{n^f}|
\tilde{q}^{{1\over 2}(L_0+\tilde{L}_0-{D\over 12})}
|B_{n^f}\rangle_u\nonumber\\
&&~=(2\pi\delta(0))^{-D}(\det(2G))^{1\over 2}\,
\eta(-1/\tilde{\tau})^{-D}
\sum_m e^{-{2\pi i\over\tilde{\tau}}mGm}\,,
\end{eqnarray}
where we have used eqs.~(\ref{eq:Bnf}) and (\ref{eq:normD}).
{}From (\ref{eq:BBt}) and (\ref{eq:BBu}), the coefficient $c_t$ 
(\ref{eq:ct}) is evaluated as
\begin{eqnarray}
\label{eq:ct_derive}
 c_t&=&\sqrt{{\cal C}_D\over {\cal C}_D'}
=\lim_{\tilde{\tau}\rightarrow +i0}\left[
{(\det(2G))^{1\over 2}\over (2\pi\delta(0))^{D}}\,
{\vartheta_2(0|-1/\tilde{\tau})^{D\over 2}\over 
\eta(-1/\tilde{\tau})^{3D\over 2}}
\sum_m e^{-{2\pi i\over\tilde{\tau}}mGm}
\right]^{1\over 2}\nonumber\\
&=&2^{D\over 4}(\det(2G))^{1\over 4}(2\pi\delta(0))^{-{D\over 2}}\,.
\end{eqnarray}
We have used eqs.~(\ref{eq:BBt}) and (\ref{eq:BBu}) instead of ${\cal
C}_D',{\cal C}_D$, respectively. Although this replacement itself is
valid up to factor, their ratio ${\cal C}_D'/{\cal C}_D$ is invariant
because they are related by the same conformal mapping 
(\ref{eq:Mandel1loop}).

In the case of Neumann type boundary states, we evaluate $c_t$ in the
same way as above.
In the twisted sector
$({\cal C}_D')$, we can use the same value as the Dirichlet type
(\ref{eq:BBt})  because of the identity:
\begin{eqnarray}
\label{eq:BBtN}
{}_t\langle B_{m^f},F|\tilde{q}^{{1\over 2}(L_0+\tilde{L}_0-2a_t)}
|B_{m^f},F\rangle_t
={}_t\langle B_{n^f}|\tilde{q}^{{1\over 2}(L_0+\tilde{L}_0-2a_t)}
|B_{n^f}\rangle_t\,,
\end{eqnarray}
which follows from (\ref{eq:Bmft}).
On the other hand, for untwisted sector, we replace (\ref{eq:BBu})
with
\begin{eqnarray}
\label{eq:BBuN}
&& (2\pi\delta(0))^{-D}{\det}^{1\over 2}(2G^{-1}_O)\,{}_u\langle B_{m^f},F|
\tilde{q}^{{1\over 2}(L_0+\tilde{L}_0-{D\over 12})}
|B_{m^f},F\rangle_u\nonumber\\
&&~=(2\pi\delta(0))^{-D}(\det(2G_O^{-1}))^{1\over 2}\,
\eta(-1/\tilde{\tau})^{-D}
\sum_{m}
e^{-{2\pi i\over \tilde{\tau}}m_iG_O^{ij}m_j},
\end{eqnarray}
to evaluate ${\cal C}_D$. Note (\ref{eq:uu_N}) and
(\ref{eq:Bmf})
comparing to (\ref{eq:CDdef})
for the prefactor. We have used the modular transformation 
in (\ref{eq:normN}).
This gives the ratio of the determinant $c_t=
2^{D\over 4}(\det(2G^{-1}_O))^{1\over 4}(2\pi\delta(0))^{-{D\over 2}}$
and the coefficient of the twisted term of
 (\ref{eq:fractional_DN}), which is 
consistent with T-duality transformation: $G\rightarrow G^{-1}_O$
compared to Dirichlet type idempotents~(\ref{eq:fractional_D}).

\section{Some formulae
\label{sec:formulae}
}

For the operators $a_i,a^{\dagger}_j$ such as
$[a_i,a^{\dagger}_j]=\delta_{ij}$ and 
$[a_i,a_j]=[a^{\dagger}_i,a^{\dagger}_j]=0$, we have a normal ordering
formula:
\begin{eqnarray}
\label{eq:KP}
&&e^{a^{\dagger}Aa^{\dagger}+aBa+{1\over 2}(a^{\dagger}Ca
+aC^Ta^{\dagger})+Da^{\dagger}+Ea}\nonumber \\
&&=~\det{}^{-{1\over 2}}(1-C)\,e^{E{4-C\over 12(1-C)}AE^T
+DB{4-C\over 12(1-C)}D^T+E{6-4C+C^2\over 12(1-C)}D^T}\\
&&~~~~\times\, e^{a^{\dagger}(1-C)^{-1}Aa^{\dagger}
+\left(EA+D\left(1-{C^T\over 2}\right)\right)(1-C^T)^{-1}a^{\dagger}}
e^{-a^{\dagger}\log (1-C) a}
e^{aB(1-C)^{-1}a
+\left(E\left(1-{C\over 2}\right)+DB\right)(1-C)^{-1}a}~~~~
\nonumber
\end{eqnarray}
for matrices $A,B,C$, which satisfy the relations
\begin{equation}
\label{eqn:ABC}
A^T=A,\ \ B^T=B,\ \ C^2=4AB,\ \ AC^T=CA,\ \ C^TB=BC,
\end{equation}
and vectors $D,E$.
This formula is obtained, for example,  by using similar technique 
in \cite{KP} Appendix A.

We use following formulae in order to compute (\ref{eq:VDstar})
explicitly:
\begin{eqnarray}
\sum_{n=1}^{\infty}
{\sin nx\over n}&=&-{1\over 2}x+{1\over 2}\pi\epsilon(x)\,,
~~~~~(|x|\le 2\pi)\,,\nonumber\\
\sum_{n=1}^{\infty}{\sin nx\sin ny\over n^2}&=&
{x(\pi-y)\over 2}-{\pi(x-y)\over 2}\theta(x-y)\,,
~~~~(-y\le x\le 2\pi-y)\,,
\label{eq:formulae}
\end{eqnarray}
where $\epsilon(x),\theta(x)$ are sign and step function respectively.


\end{document}